\documentclass[12pt,a4paper]{article}
\usepackage{myart}
\usepackage{cite}
\usepackage{amsmath}
\usepackage{graphicx}
\usepackage{amsfonts}
\usepackage{amssymb}

\begin{document}
\hfill\hbox{LPM-00-15}

\hfill\hbox{May, 2000}

\bigskip\ 

\begin{center}
{\Large \textbf{On the Thermodynamic Bethe Ansatz Equation in Sinh-Gordon
Model}}

\vspace{1.5cm}

{\large Al.Zamolodchikov}\footnote{On leave of absence from Institute of
Theoretical and Experimental Physics, B.Cheremushkinskaya 25, 117259 Moscow, Russia}

\vspace{0.2cm}

Laboratoire de Physique Math\'ematique\footnote{Laboratoire Associ\'e au CNRS URA-768}

Universit\'e Montpellier II

Pl.E.Bataillon, 34095 Montpellier, France
\end{center}

\vspace{1.0cm}

\textbf{Abstract}

Two implicit periodic structures in the solution of sinh-Gordon thermodynamic
Bethe ansatz equation are considered. The analytic structure of the solution
as a function of complex $\theta$ is studied to some extent both analytically
and numerically. The results make a hint how the CFT integrable structures can
be relevant in the sinh-Gordon and staircase models. More motivations are
figured out for subsequent studies of the massless sinh-Gordon (i.e.
Liouville) TBA equation.

\section{Sinh-Gordon model\label{ShG}}

It seems that from the recent developments of the string theory there are some
persistent requests about better understanding of the two-dimensional
sigma-models with non-compact (in particular singular) target spaces. Physical
properties of such models are expected to be quite different from these of the
better studied compact sigma-models. It is therefore a challenge for the
two-dimensional integrable field theory community to reveal the corresponding
peculiarities and new features.

The two-dimensional sinh-Gordon model (ShG) is defined by the (Euclidean)
action
\begin{equation}
A_{\mathrm{ShG}}=\int\left[  \frac1{4\pi}(\partial_{a}\phi)^{2}+2\mu
\cosh(2b\phi)\right]  d^{2}x\label{shg}%
\end{equation}
I believe that this model can be considered to say as ``a model'' of (suitably
perturbed) non-compact sigma-models, in the sense that its properties are
considerably different from those of the perturbed rational conformal field
theories, these differences being sometimes quite similar to these between
non-compact and compact sigma-models.

In (\ref{shg}) $\mu$ is a dimensional (of dimension $\mu\sim[\mathrm{mass}%
]^{2+2b^{2}}$) coupling constant which determines the scale in the model and
$b $ is the dimensionless ShG parameter. For the beginning I suppose it to be
real and non-negative. The case $b=0$ turns to be somewhat singular for the
subsequent considerations and here the study is restricted to positive values
$0<b<\infty$ only. It is also convenient to use different parameters
\begin{equation}
Q=b+1/b\label{Qb}%
\end{equation}
and
\begin{equation}
p=\frac bQ=\frac{b^{2}}{1+b^{2}}\label{p}%
\end{equation}
instead of $b$.

The perturbing operators $\exp(\pm2b\phi)$ in (\ref{shg}) have negative
dimension $\Delta=-b^{2}$. To make the coupling $\mu$ a strict sense we need
to fix a normalization of these operators. Here they are implied to be normal
ordered w.r.t. the massless (unperturbed) vacuum in the way that in
unperturbed theory
\begin{equation}
\left\langle e^{2b\varphi(x_{1})}\ldots e^{2b\varphi(x_{n})}e^{-2b\varphi
(y_{1})}\ldots e^{-2b\varphi(y_{n})}\right\rangle _{\mu=0}=\frac{\prod
_{i,j}|x_{i}-y_{j}|^{4b^{2}}}{\prod_{i>j}(|x_{i}-x_{j}||y_{i}-y_{j}|)^{4b^{2}%
}}\label{coulomb}%
\end{equation}

The model is known to be integrable and can be solved for many important
characteristics. In particular its factorized scattering theory is known since
long \cite{Gryanik, Korepin}. The spectrum consists of only one neutral
particle $B(\theta)$ subject to a factorized scattering with two-particle
amplitude
\begin{equation}
S(\theta)=\frac{\sinh\theta-i\sin\pi p}{\sinh\theta+i\sin\pi p}\label{S}%
\end{equation}
With the normalization (\ref{coulomb}) the mass $m$ of this particle is
related to the scale parameter $\mu$ as \cite{mscale}
\begin{equation}
\pi\mu\frac{\Gamma(b^{2})}{\Gamma(1-b^{2})}=\left[  mZ(p)\right]  ^{2+2b^{2}%
}\label{mum}%
\end{equation}
where
\begin{equation}
Z(p)=\frac1{8\sqrt{\pi}}p^{p}(1-p)^{1-p}\Gamma\left(  \frac{1-p}2\right)
\Gamma\left(  \frac p2\right) \label{Zp}%
\end{equation}

Notice that the scattering theory is invariant under the (week-strong
coupling) duality transformation $b\rightarrow1/b$ which brings $p\rightarrow
1-p$. This means that the physical content of the model remains unchanged up
to the overall mass scale. Since the combination (\ref{Zp}) is again invariant
under $p\rightarrow1-p$, the mass scale also remains unchanged if the coupling
constant $\mu$ is simultaneously substituted by the ``dual'' coupling constant
$\tilde\mu$ related to $\mu$ as follows
\begin{equation}
\left(  \pi\mu\frac{\Gamma(b^{2})}{\Gamma(1-b^{2})}\right)  ^{1/b}=\left(
\pi\tilde\mu\frac{\Gamma(1/b^{2})}{\Gamma(1-1/b^{2})}\right)  ^{b}%
\label{mutwiddle}%
\end{equation}
Therefore the sinh-Gordon model is completely invariant under duality
$b\rightarrow1/b$; $\mu\rightarrow\tilde\mu$. Due to this symmetry it is
sufficient to consider only the region $0<b^{2}\leq1$ or $0<p\leq1/2$.

The infinite volume bulk vacuum energy of the model is also known exactly
\cite{deVega}. In terms of the particle mass $m$ it is given by the following
apparently self-dual expression
\begin{equation}
\mathcal{E}=\frac{m^{2}}{8\sinh\pi p}\label{Evac}%
\end{equation}

\section{TBA equation\label{TBAsect}}

Contrary to the on-mass-shell data of ShG quoted above, the off-mass-shell
characteristics such as the correlation functions (with the exception of the
vacuum expectation values of some local fields \cite{onepoint, descendents}
and their matrix elements between the asymptotic states \cite{simonetti}) are
not known exactly.

Some progress can be made for the finite size effects where the problem is
reduced to a non-linear integral equation known as the thermodynamic Bethe
ansatz (TBA) one. Namely, consider the ground state energy $E(R)$ of the
finite size ShG model placed on a circle of finite circumference $R$. In the
TBA framework it appears as
\begin{equation}
E(R)=-\frac m{2\pi}\int\cosh\theta\log\left(  1+e^{-\varepsilon(\theta
)}\right)  d\theta\label{E0}%
\end{equation}
It is convenient to introduce also the $R$-dependent effective central charge
$c_{\text{eff}}(R)$ as
\begin{equation}
c_{\text{eff}}(R)=-\frac{6R}\pi E(R)\label{ceff}%
\end{equation}
In (\ref{E0}) $\varepsilon(\theta)$ is the solution of the TBA equation
\begin{equation}
mR\cosh\theta=\varepsilon+\varphi*\log\left(  1+e^{-\varepsilon(\theta
)}\right) \label{TBA}%
\end{equation}
where $*$ denotes the $\theta$-convolution. The kernel $\varphi(\theta)$ is
related to the ShG scattering data (\ref{S})
\begin{equation}
\varphi(\theta)=-\frac i{2\pi}\frac d{d\theta}\log S(\theta)=\frac1{2\pi}%
\frac{4\sin\pi p\cosh\theta}{\cosh2\theta-\cos2\pi p}\label{kernel}%
\end{equation}
The Fourier transform of the kernel reads
\begin{equation}
\varphi(\omega)=\int e^{i\omega\theta}\varphi(\theta)d\theta=\frac{\cosh
\dfrac{a\pi\omega}2}{\cosh\dfrac{\pi\omega}2}\label{Fourier}%
\end{equation}
where the parameter
\begin{equation}
a=1-2p\label{a}%
\end{equation}
is simply reflected $a\rightarrow-a$ under the dualily $p\rightarrow1-p$ and
therefore can be taken non-negative $0\leq a<1$.

The following conclusion are readily made from the structure of the integral
equation (\ref{TBA}).

\textbf{1.} Function is even $\varepsilon(\theta)=\varepsilon(-\theta)$ and
analytic in the strip $|\operatorname*{Im}\theta|<\pi/2-\pi a/2$. At
$\operatorname*{Re}\theta\rightarrow\infty$ in this strip it has the
asymptotic $\varepsilon(\theta)\sim mRe^{\theta}/2$. Therefore the function
\begin{equation}
Y(\theta)=\exp(-\varepsilon(\theta))\label{Y}%
\end{equation}
is analyic and non-zero in this strip and at $\operatorname*{Re}%
\theta\rightarrow\infty$ behaves as
\begin{equation}
Y(\theta)\sim\exp(-mRe^{\theta}/2)\label{Yasymp}%
\end{equation}
The asymptotic at $\mathrm{\operatorname*{Re}}\theta\rightarrow-\infty$ is
related to (\ref{Yasymp}) by the symmetry $Y(\theta)=Y(-\theta)$. Let us
define another even function
\begin{equation}
X(\theta)=\exp\left[  -\frac{mR}{2\sin\pi p}\cosh\theta+\int\frac
{\log(1+Y(\theta^{\prime}))}{\cosh(\theta-\theta^{\prime})}\frac
{d\theta^{\prime}}{2\pi}\right] \label{X}%
\end{equation}
which is obviously analytic and non-zero in the strip $|\operatorname*{Im}%
\theta|<\pi/2$ and at $\operatorname*{Re}\theta\rightarrow\infty$ in this
strip
\begin{equation}
X(\theta)\sim\exp\left(  -\frac{mR}{4\sin\pi p}\exp\theta\right)
\label{Xasymp}%
\end{equation}
From the TBA equation it follows that
\begin{equation}
X\left(  \theta+ia\pi/2\right)  X\left(  \theta-ia\pi/2\right)  =Y(\theta
)\label{XXY}%
\end{equation}

\textbf{2.} In fact on the real axis $Y(\theta)$ is real and positive and
therefore we expect a strip $|\operatorname*{Im}\theta|<\epsilon$ with some
finite $\epsilon>0$ where $1+Y(\theta)\neq0$. Therefore the analyticity
condition for $X(\theta)$ can be extended to the strip $|\operatorname*{Im}%
\theta|<\pi/2+\epsilon$. This is enough to prove the relation
\begin{equation}
X\left(  \theta+i\pi/2\right)  X\left(  \theta-i\pi/2\right)  =1+Y(\theta
)\label{XX1Y}%
\end{equation}
The functional equation
\begin{equation}
X\left(  \theta+i\pi/2\right)  X\left(  \theta-i\pi/2\right)  =1+X\left(
\theta+ia\pi/2\right)  X\left(  \theta-ia\pi/2\right) \label{XX}%
\end{equation}
follows. This relation allows to extend the original analyticity strip
$|\operatorname*{Im}\theta|<\pi/2+\epsilon$ to the strip $|\operatorname*{Im}%
\theta|<3\pi/2$ and, as we'll see before long, to the whole comlex plane of
$\theta$ so that $X(\theta)$ is an entire function of $\theta$. Notice that
from (\ref{XX}) it follows that the asymptotic (\ref{Xasymp}) holds in the
larger strip $|\operatorname*{Im}\theta|<\pi$. The asymptotic outside this
strip is more complicated.

\textbf{3.} As a consequence, $Y(\theta)$ is also entire function of $\theta$
and satisfyes the following functional relation
\begin{equation}
Y\left(  \theta+i\pi/2\right)  Y\left(  \theta-i\pi/2\right)  =\left(
1+Y\left(  \theta+ia\pi/2\right)  \right)  \left(  1+Y\left(  \theta
-ia\pi/2\right)  \right) \label{YY}%
\end{equation}

The last equation is very similar to the functional relations appearing in the
TBA study of the integrable perturbed rational conformal field theories (the
so called $Y$-systems). Typically such $Y$-systems imply a periodicity of the
$Y$-functions in $\theta$ with some imaginary period related to the scale
dimension $\Delta$ of the perturbing operator (see e.g., \cite{Zam1,
Ravanini}). This periodicity in order entails special ``perturbative''
structure of the short distance $R\rightarrow0$ behavior of the ground state
energy $E(R)$. Namely, up to one exceptional term, it is a regular expansion
in powers of $R^{2-2\Delta}$%
\begin{equation}
E(R)=-\mathcal{E}_{\mathrm{vac}}R-\frac\pi{6R}\sum_{n=0}^{\infty}%
c_{n}R^{(2-2\Delta)n}\;;\;\;\;\;R\rightarrow0\label{Ereg}%
\end{equation}
where $\mathcal{E}_{\mathrm{vac}}$ is the infinite volume vacuum energy of the
model. Unlike this typical situation, the $Y$-system (\ref{YY}) or the $X
$-system (\ref{XX}) does not imply any apparent periodic structure of
$Y(\theta)$ in $\theta$. As a manifestation of this peculiarity, the
$R\rightarrow0$ behavior of $E(R)$ is different from (\ref{Ereg}) and includes
softer logarithmic corrections \cite{staircase}
\begin{equation}
E(R)=-\mathcal{E}R-\frac\pi{6R}\left(  1-\frac{3\pi^{2}p(1-p)}{2\log^{2}%
R}+O\left(  \log^{-3}R\right)  \right)  \;;\;\;\;\;R\rightarrow0\label{Eirreg}%
\end{equation}
The purpose of the next two sections is to reveal two hidden periodic
structures (with different periods) of the $Y$-system (\ref{YY}).

\section{Discrete Liouville equation\label{DLE}}

In this section I discuss the following two-dimensional non-linear
finite-difference equation for the function $X(u,v)$%
\begin{equation}
X(u+1,v)X(u-1,v)=1+X(u,v+1)X(u,v-1)\label{XXuv}%
\end{equation}
which is apparently resemblent of the functional $X$-system (\ref{XX}). In the
next section we'll see how some of the results for (\ref{XXuv}) can be
specialized to our TBA problem. Equation (\ref{XXuv}) is a particular case of
Hirota difference equation \cite{Hirota}. The constructions of this section
can be found in \cite{Wiegmann} (see also \cite{Zabrodin}) where more general
difference system is analysed. They appeared also in a quite close context in
\cite{BLZ2}.

Equation (\ref{XXuv}) can be considered as a discretisation of the hyperbolic
Liouville equation
\begin{equation}
\partial_{u}^{2}\varphi-\partial_{v}^{2}\varphi=-e^{2\varphi}\label{Liouville}%
\end{equation}
Indeed, let $X(u,v)=\exp(-\varphi(u,v))$ and let $\varphi(u,v)$ be large and
negative. Then eq.(\ref{XXuv}) is approximated by
\begin{equation}
\varphi(u+1,v)+\varphi(u-1,v)-\varphi(u,v+1)-\varphi(u,v-1)=-\exp
(\varphi(u,v+1)+\varphi(u,v-1))\label{phixy}%
\end{equation}
In the long-wave limit this is reduced to (\ref{Liouville}). As we'll see
below, eq.(\ref{XXuv}) is in many respects very similar to the Liouville
equation. It seems quite natural to call it the discrete Liouville equation
\cite{Hirota1}.

Let me remind well known construction of a local solution to
eq.(\ref{Liouville}).

\textbf{1a.} It is convenient to use the light cone variables $x^{+}=u+v$ and
$x^{-}=u-v$, so that $\partial_{+}=(\partial_{u}+\partial_{v})/2$;
$\partial_{-}=(\partial_{u}-\partial_{v})/2$ and (\ref{Liouville}) reads
\begin{equation}
4\partial_{+}\partial_{-}\varphi=-e^{2\varphi}\label{Luv}%
\end{equation}
Let $\varphi$ be a local solution of (\ref{Luv}). Define
\begin{align}
t  & =-(\partial_{+}\varphi)^{2}+\partial_{+}^{2}\varphi\label{T}\\
\tilde t  & =-(\partial_{-}\varphi)^{2}-\partial_{-}^{2}\varphi\nonumber
\end{align}
As a consequence of eq.(\ref{Luv}) we have
\begin{equation}
\partial_{-}t=\partial_{+}\tilde t=0\label{dT}%
\end{equation}
so that $t(x^{+})$ and $\tilde t(x^{-})$ are respectively functions of only
$x^{+}$ and $x^{-}$.

\textbf{2a.} Field $X=\exp(-\varphi)$ satisfies two linear differential
equations
\begin{equation}
(\partial_{+}^{2}+t(x^{+}))X(u,v)=0\;;\;\;\;(\partial_{-}^{2}+\tilde
t(x^{-}))X(u,v)=0\label{d2X}%
\end{equation}

\textbf{3a.} Let $Q_{\pm}(x)$ and $\tilde Q_{\pm}(x)$ be linearly independent
solutions to the ordinary differential equations
\begin{align}
(\partial_{x}^{2}+t(x))Q_{\pm}(x)  & =0\label{dQ}\\
(\partial_{x}^{2}+\tilde t(x))\tilde Q_{\pm}(x)  & =0\nonumber
\end{align}
normalized in the way that
\begin{align}
\partial_{x}Q_{+}(x)Q_{-}(x)-Q_{+}(x)\partial_{x}Q_{-}(x)  & =1/2\label{W}\\
\partial_{x}\tilde Q_{+}(x)\tilde Q_{-}(x)-\tilde Q_{+}(x)\partial_{x}\tilde
Q_{-}(x)  & =1/2\nonumber
\end{align}
A local solution of (\ref{Luv}) can be constructed as
\begin{equation}
\exp(-\varphi(u,v))=Q_{+}(x^{+})\tilde Q_{+}(x^{-})+Q_{-}(x^{+})\tilde
Q_{-}(x^{-})\label{phiQQ}%
\end{equation}

\textbf{4a.} Introduce the functions
\begin{equation}
F(x)=\frac{Q_{+}(x)}{Q_{-}(x)}\;;\;\;\;G(x)=-\frac{\tilde Q_{-}(x)}{\tilde
Q_{+}(x)}\label{FG}%
\end{equation}
The solution (\ref{phiQQ}) can be rewritten as
\begin{equation}
\exp2\varphi(u,v)=\frac{4F^{\prime}(x^{+})G^{\prime}(x^{-})}{(F(x^{+}%
)-G(x^{-}))^{2}}\label{phiFG}%
\end{equation}
Notice also that in terms of $F$ and $G$%
\begin{equation}
t(x)=-2\{F(x),x\}\;;\;\;\;\tilde t(x)=-2\{G(x),x\}\label{TFG}%
\end{equation}
where
\begin{equation}
\{f(x),x\}=\frac{f^{\prime\prime\prime}}{f^{\prime}}-\frac32\left(
\frac{f^{\prime\prime}}{f^{\prime}}\right)  ^{2}\label{Schwarz}%
\end{equation}
is the Schwarz derivative.

Now let us turn to the discrete Liouville equation (\ref{XXuv}). Defining,
similarly to eq.(\ref{XXY}),
\begin{align}
Y(u,v)  & =X(u,v+1)X(u,v-1)\label{split}\\
1+Y(u,v)  & =X(u+1,v)X(u-1,v)\nonumber
\end{align}
we arrive at the finite difference equation analogous to (\ref{YY})
\begin{equation}
Y(u+1,v)Y(u-1,v)-(1+Y(u,v+1))(1+Y(u,v-1))\label{YYuv}%
\end{equation}

\textbf{1b.} Introduce
\begin{align}
T(u,v)  & =\frac{X(u+1,v+1)+X(u-1,v-1)}{X(u,v)}\label{TT}\\
\tilde T(u,v)  & =\frac{X(u+1,v-1)+X(u-1,v+1)}{X(u,v)}\nonumber
\end{align}
As a consequence of eq.(\ref{XXuv}) we obtain
\begin{align}
T(u+1,v-1)  & =T(u,v)\label{DT}\\
\tilde T(u+1,v+1)  & =\tilde T(u,v)\nonumber
\end{align}
so that $T=T(u+v)$ and $\tilde T=\tilde T(u-v)$. In the continuous limit these
objects are related to (\ref{T}) as $T(u)=2-4t(u)+\ldots$; $\tilde
T(u)=2-4\tilde t(u)+\ldots$ \cite{BLZ2}.

\textbf{2b.} Eqs.(\ref{TT}) can be rewritten as (similarly to (\ref{d2X}))
\begin{align}
X(u+1,v+1)+X(u-1,v-1)  & =T(u+v)X(u,v)\label{D2X}\\
X(u+1,v-1)+X(u-1,v+1)  & =\tilde T(u-v)X(u,v)\nonumber
\end{align}

\textbf{3b.} Let $Q_{\pm}(u)$ and $\tilde Q_{\pm}(u)$ be linearly independent
solutions of the second order finite difference equations
\begin{align}
Q_{\pm}(u+2)+Q_{\pm}(u-2) &  =T(u)Q_{\pm}(u)\label{TQ}\\
\tilde Q_{\pm}(u+2)+\tilde Q_{\pm}(u-2) &  =\tilde T(u)\tilde Q_{\pm
}(u)\nonumber
\end{align}
normalized by the ``quantum Wronskians''
\begin{align}
Q_{+}(u+1)Q_{-}(u-1)-Q_{+}(u-1)Q_{-}(u+1) &  =1\label{QW}\\
\tilde Q_{+}(u+1)\tilde Q_{-}(u-1)-\tilde Q_{+}(u-1)\tilde Q_{-}(u+1) &
=1\nonumber
\end{align}
Then it is verified that
\begin{equation}
X(u,v)=Q_{+}(u+v)\tilde Q_{+}(u-v)+Q_{-}(u+v)\tilde Q_{-}(u-v)\label{XQQ}%
\end{equation}
is a local solution of the discrete Liouville equaiton (\ref{XXuv}).

\textbf{4b.} Introduce the functions
\begin{equation}
F(u)=\frac{Q_{+}(u)}{Q_{-}(u)}\;;\;\;\;G(u)=-\frac{\tilde Q_{-}(u)}{\tilde
Q_{+}(u)}\label{QFG}%
\end{equation}
which can be used to present the local solution (\ref{XQQ}) in the form
\begin{align}
Y(u,v)  & =\frac{(F(u+v+1)-G(u-v-1))(F(u+v-1)-G(u-v+1))}%
{(F(u+v+1)-F(u+v-1))(G(u-v+1)-G(u-v-1))}\label{YFG}\\
1+Y(u,v)  & =\frac{(F(u+v+1)-G(u-v+1))(F(u+v-1)-G(u-v-1))}%
{(F(u+v+1)-F(u+v-1))(G(u-v+1)-G(u-v-1))}\nonumber
\end{align}
Let me mention also the discrete analogue of the Schwarzian derivative
(\ref{TFG})
\begin{equation}
T(u+1)T(u-1)=\frac{(F(u+3)-F(u-1))(F(u+1)-F(u-3))}%
{(F(u+3)-F(u+1))(F(u-1)-F(u-3))}\label{DS}%
\end{equation}

\section{Application to TBA\label{Application}}

The above constructions for the discrete Liouville equation can be translated
for the ShG $X$-system (\ref{XX}) we are interested in. Let us require the
following periodicity condition for $X(u,v)$ in (\ref{XXuv})
\begin{equation}
X(u+a,v+1)=X(u,v)\label{Xpbc}%
\end{equation}
with some parameter $a$ (at this point we start to diverge from the lines of
\cite{BLZ2}). With this periodicity eq.(\ref{XXuv}) reads
\begin{equation}
X(u+1,v)X(u-1,v)=1+X(u+a,v)X(u-a,v)\label{XXa}%
\end{equation}
Here $v$ can be considered as a parameter. Suppressing this redundant
dependence and rescaling $u$ as
\begin{equation}
\theta=i\pi u/2\label{thetau}%
\end{equation}
we are back to the ShG $X$ system in the form (\ref{XX}).

From $X(\theta)$ the two functions $T(\theta)$ and $\tilde T(\theta)$ are
readily restored as
\begin{align}
T(\theta) &  =\frac{X\left(  \theta+i\pi(1-a)/2\right)  +X\left(  \theta
-i\pi(1-a)/2\right)  }{X(\theta)}\label{shGT}\\
\tilde T(\theta) &  =\frac{X\left(  \theta+i\pi(1+a)/2\right)  +X\left(
\theta-i\pi(1+a)/2\right)  }{X(\theta)}\nonumber
\end{align}
The ``holomorphic'' property (\ref{DT}) is translated to the following
periodicity of these functions
\begin{align}
T\left(  \theta+i\pi(1+a)/2\right)   &  =T(\theta)\label{periodicity}\\
\tilde T\left(  \theta+i\pi(1-a)/2\right)   &  =\tilde T(\theta)\nonumber
\end{align}
Notice that the period $i\pi/(1+b^{2})$ of $T$ corresponds to the negative
dimension $\Delta=-b^{2}$ of the perturbing operator in (\ref{shg}). As it can
be anticipated from the self-duality of ShG, the second period $i\pi
b^{2}/(1+b^{2})$ of $\tilde T$ is related to the dimension $\tilde
\Delta=-b^{-2}$ of the ``dual'' exponentials $\exp(\pm2\phi/b)$.

As it is discussed in sect.2 $X(\theta)$ is analytic and non-zero in the strip
$|\operatorname*{Im}\theta|<\pi/2$ and analytic in the larger strip
$|\operatorname*{Im}\theta|<3\pi/2$. Therefore $T(\theta)$ and $\tilde
T(\theta)$ are analytic in the strip $|\operatorname*{Im}\theta|<\pi/2$ and by
periodicity (\ref{periodicity}) are entire functions of $\theta$. It follows
from (\ref{shGT}) that $X(\theta)$ is an entire function of $\theta$ too.

The asymptotics at $\operatorname*{Re}\theta\rightarrow\infty$ follow from
(\ref{Xasymp}) and (\ref{shGT})
\begin{align}
T(\theta) &  \sim\exp\left(  \frac{mR\exp(\theta-i\pi(1-p)/2)}{4\cos(\pi
p/2)}\right)  \;\;\;\;\text{in the strip }0<\operatorname*{Im}\theta
<\pi(1+a)/2\label{Tasymp}\\
\tilde T(\theta) &  \sim\exp\left(  \frac{mR\exp(\theta-i\pi p/2)}{4\sin(\pi
p/2)}\right)  \;\;\;\;\;\;\;\;\;\;\;\;\;\text{in the strip }%
0<\operatorname*{Im}\theta<\pi(1-a)/2\nonumber
\end{align}
The real axis $\operatorname*{Im}\theta=0$ is a Stokes line and here
\begin{align}
T(\theta) &  \sim2\exp\left(  \frac{mR}4\tan(\pi p/2)e^{\theta}\right)
\cos\left(  \frac{mR}4\exp\theta\right) \label{Tstokes}\\
\tilde T(\theta) &  \sim2\exp\left(  \frac{mR}4\cot(\pi p/2)e^{\theta}\right)
\cos\left(  \frac{mR}4\exp\theta\right) \nonumber
\end{align}
The $\operatorname*{Re}\theta\rightarrow\infty$ asymptotic in the whole plane
of $\theta$ is restored from the periodicity (\ref{periodicity}). Following
(\ref{Tstokes}) both $T$ and $\tilde T$ have infinite number of zeroes on the
real axis located at $\theta=\pm\theta_{n}$, $n=1,2,\ldots,\infty$ with
$\theta_{n}\sim\log(2\pi n/mR)+O(1/n)$ at $n\rightarrow\infty$. The
half-period shifted functions $T(\theta+i\pi(1-p)/2)$ and $\tilde
T(\theta+i\pi p/2)$ are also real at real $\theta$ and at $\theta
\rightarrow\infty$ behave as
\begin{align}
T(\theta+i\pi(1-p)/2) &  \sim\exp\left(  \frac{mR}{4\cos(\pi p/2)}e^{\theta
}\right) \label{Thalf}\\
\tilde T(\theta+i\pi p/2) &  \sim\exp\left(  \frac{mR}{4\sin(\pi
p/2)}e^{\theta}\right) \nonumber
\end{align}

Analytic properties of $T$ and $\tilde T$ allow the following convergent
expansions
\begin{align}
T(\theta) &  =\sum_{-\infty}^{\infty}T_{n}\exp(2nQb\theta)\label{Tch}\\
\tilde T(\theta) &  =\sum_{-\infty}^{\infty}\tilde T_{n}\exp(2nQ\theta
/b)\nonumber
\end{align}
with real $T_{n}$ and $\tilde T_{n}$. From the symmetry of the original TBA
equations $T$ and $\tilde T$ both even functions of $\theta$ so that in our
present case $T_{n}=T_{-n}$ and $\tilde T_{n}=\tilde T_{-n}$. One can read-off
the leading large $n$ behavior of the coefficients $T_{n}$ and $\tilde T_{n}$
from the asymptotics (\ref{Tasymp})
\begin{align}
T_{n}  & \sim(-)^{n}\sqrt{\frac{bQ}\pi}n^{-2Qbn-1/2}\left(  e\frac{mR}{8\pi
bQ\cos(\pi p/2)}\right)  ^{2Qbn}\label{Tn}\\
\tilde T_{n}  & \sim(-)^{n}\sqrt{\frac Q{\pi b}}n^{-2Qn/b-1/2}\left(
e\frac{bmR}{8\pi Q\sin(\pi p/2)}\right)  ^{2Qn/b}\nonumber
\end{align}

So far the constructions were explicitely based on the integral equation
(\ref{TBA}). The rest of the section is more speculative. Suppose that for
$T(\theta)$ and $\tilde T(\theta)$ constructed as in (\ref{shGT}) we can find
$Q_{\pm}(\theta)$ and $\tilde Q_{\pm}(\theta)$ which solve
\begin{align}
Q_{\pm}\left(  \theta+i\pi\right)  +Q_{\pm}(\theta-i\pi) &  =T(\theta)Q_{\pm
}(\theta)\label{QQTQ}\\
\tilde Q_{\pm}\left(  \theta+i\pi\right)  +\tilde Q_{\pm}(\theta-i\pi) &
=T(\theta)\tilde Q_{\pm}(\theta)\nonumber
\end{align}
and are the ``Bloch waves'' with respect to the periods of $T$ and $\tilde T$
respectively
\begin{align}
Q_{\pm}\left(  \theta+i\pi(1+a)/2\right)   &  =\exp\left(  \pm2i\pi
P/Q\right)  Q_{\pm}(\theta)\label{Bloch}\\
\tilde Q_{\pm}\left(  \theta+i\pi(1-a)/2\right)   &  =\exp\left(  \pm2i\pi
P/Q\right)  \tilde Q_{\pm}(\theta)\nonumber
\end{align}
with some Floquet index $P$. Let them be normalized by the quantum Wronskians
\begin{align}
Q_{+}(\theta+i\pi/2)Q_{-}(\theta-i\pi/2)-Q_{+}(\theta-i\pi/2)Q_{-}(\theta
+i\pi/2) &  =1\label{qW1}\\
\tilde Q_{+}(\theta+i\pi/2)\tilde Q_{-}(\theta-i\pi/2)-\tilde Q_{+}%
(\theta-i\pi/2)\tilde Q_{-}(\theta+i\pi/2) &  =1\nonumber
\end{align}
Then formally
\begin{equation}
X(\theta)=Q_{+}(\theta)\tilde Q_{+}(\theta)+Q_{-}(\theta)\tilde Q_{-}%
(\theta)\label{XQQ1}%
\end{equation}
solves eqs.(\ref{shGT}) as well as the $X$-system (\ref{XX}).

Unfortunately at present I know no effective means to construct these
$Q$-functions. Moreover, there are serious doubts that the objects satisfying
both (\ref{QQTQ}) and (\ref{Bloch}) can be constructed in any sense, at least
at rational values of $b^{2}$. I hope to say something more definite on this
point in close future.

\section{Large $\operatorname*{Re}\theta$ asymptotics\label{LargeRe}}

Let me comment a little more about the $\operatorname*{Re}\theta
\rightarrow\infty$ asymptotics (with $\operatorname*{Im}\theta$ fixed) of the
function $X(\theta)$ in the whole complex plane. In principle it can be
restored from the asymptotic (\ref{Xasymp}) in the strip $|\operatorname*{Im}%
\theta|\leq\pi/2$ using the functional relation (\ref{XX}) or, more
conveniently, the relations (\ref{shGT}) together with the asymptotics
(\ref{Tasymp}). The asymptotics is always of the form
\begin{equation}
X(\theta)\sim\exp\left(  A(\operatorname*{Im}\theta)\exp(\operatorname*{Re}%
\theta)\right) \label{Asymp}%
\end{equation}
with some complex function $A(\eta)$ of real variable $\eta=\operatorname*{Im}%
\theta$. Apparently $\operatorname*{Re}A(\eta)$ controls the rate of growth of
the absolute value of $X$. At $|\eta|<\pi/2$ we have
\begin{equation}
A(\eta)=-\frac{mR}{4\sin\pi p}e^{i\eta}\label{Api}%
\end{equation}
It follows from (\ref{shGT}) that $A(\eta)$ satisfies two functional relations%

\begin{align}
A(\eta+\pi(1-a)/2)) &  =%
\genfrac{\{}{.}{0pt}{0}{A(\eta)+\sigma(\eta)\;\;\;\text{if \ }%
\operatorname*{Re}(A(\eta)+\sigma(\eta))>\operatorname*{Re}A(\eta
+\pi(1-a)/2)}{A(\eta+\pi(1-a)/2)\;\;\;\;\;\text{otherwise}\;}%
\label{iff}\\
A(\eta+\pi(1+a)/2)) &  =%
\genfrac{\{}{.}{0pt}{0}{A(\eta)+\tilde\sigma(\eta)\;\;\;\text{if
\ }\operatorname*{Re}(A(\eta)+\tilde\sigma(\eta))>\operatorname*{Re}A(\eta
+\pi(1+a)/2)}{A(\eta+\pi(1+a)/2)\;\;\;\;\;\text{otherwise}}%
\nonumber
\end{align}
where the functions $\sigma(\eta)$ and $\tilde\sigma(\eta)$ control the
asymptotics of $T(\theta)$ and $\tilde T(\theta)$. They are defined as
\begin{align}
\sigma(\eta) &  =\frac{mR}{4\cos(\pi p/2)}\exp(i\eta-i\pi
(1-p)/2)\;\;\;\text{at \ }0\leq\eta<(1+a)/2\label{tau}\\
\tilde\sigma(\eta) &  =\frac{mR}{4\sin(\pi p/2)}\exp(i\eta-i\pi
p/2)\;\;\;\;\;\;\;\;\;\;\;\text{at\ \ }0\leq\eta<(1-a)/2\nonumber
\end{align}
and continued outside these regions periodically as $\tilde\sigma(\eta
+\pi(1-a)/2)=\tilde\sigma(\eta)$ and $\sigma(\eta+\pi(1+a)/2)=\sigma(\eta)$.
Both $\sigma(\eta)$ and $\tilde\sigma(\eta)$ jump by $-imR/2$ at $\eta
=\pi(1+a)n/2$, $n\in Z$ and $\eta=\pi(1+a)n/2$, $n\in Z$ respectively. This
corresponds to the limiting density of zeroes prescribed by (\ref{Tstokes}).

A common solution to (\ref{iff}) exists. Generally the solution is
discontinuous at all values $\eta=\pm\pi(m(1+a)/2+n(1-a)/2)$ with arbitrary
positive integers $n$ and $m$. At each such point the imaginary part
$\operatorname*{Im}A$ jumps down by $-mR/2$ indicating an asymptotic line of
accumulation of zeroes of $X(\theta)$, the asymptotic density being the same
as that of the functions $T$ and $\tilde T$ (\ref{Tstokes}). The real part of
$A$ at these points is continuous itself but has discontinuities in the first
derivative. The first Stokes line appears at $\eta=\pi$.

At large $\eta$ the structure is qualitetively different dependent on the
arithmetic nature of parameter $b^{2}$. If it is a rational number the periods
of $T$ and $\tilde T$ are commensurable. Some of the discontinuities merge
forming multiple jumps in the imaginary part. The solution $A(\theta)$ bears a
regular ``quasiperiodic'' structure with the common period of $T$ and $\tilde
T$ . For irrational $b^{2}$ the periods are incommensurable and as
$\eta\rightarrow\infty$ the singularities are more and more dense, the
solution having quite irregular behavior.%

\begin{figure}
[ptb]
\begin{center}
\includegraphics[
height=3.4688in,
width=4.7089in
]%
{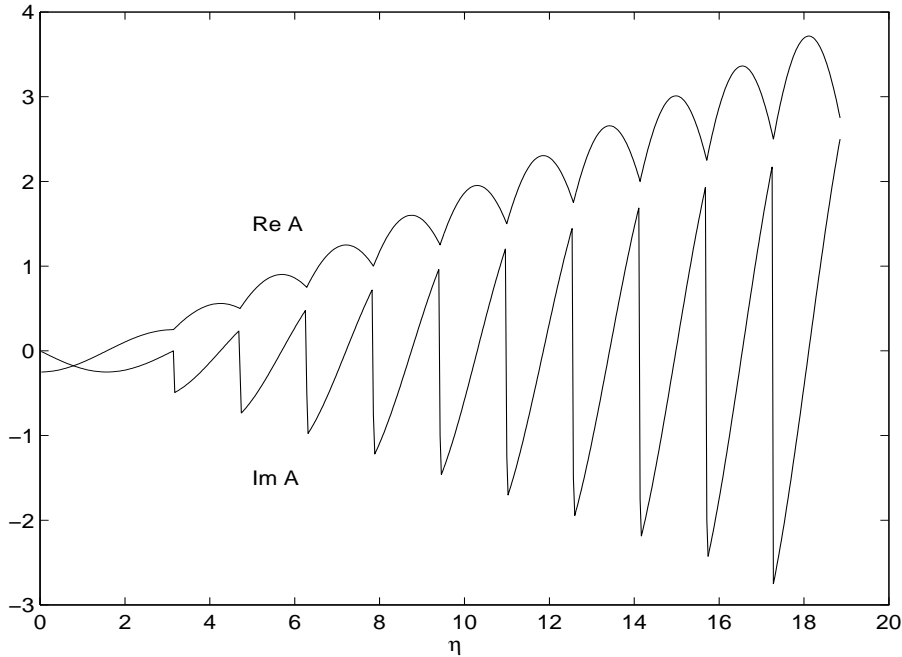}%
\caption{Real and imaginary parts of $A(\eta)/mR$ at $b^{2}=1$.}%
\label{fig1}%
\end{center}
\end{figure}

In fig.1 real and imaginary parts of $A(\eta)$ are plotted for the simplest
(self-dual) case $b=1$. Both periods are equal to $\pi/2$. The discontinuities
are located at $\eta=\pm\pi(n+1)/2$, $n=1,2,\ldots$ where the imaginary part
jumps by $-mRn/2$. Contrary to the location of zeroes of $T$, zeroes of $X$
cannot lie exactly on the lines $\mathrm{Im}\theta=$const, at least for the
first line $\eta=\pi$. Indeed, in the case $b^{2}=1$ it follows from the
functional relation that at real $\theta$%
\begin{equation}
\left|  X(\theta+i\pi)\right|  ^{2}=X^{2}(\theta)+T^{2}(\theta
)\label{Xpositive}%
\end{equation}
which is strictly positive.

Another rational situation $b^{2}=1/2$ corresponds to the periods $\pi/3$ and
$2\pi/3$. Function $A(\eta)$ is plotted in fig.2. The structure is again quite
regular, the discontinuities occuring at $\eta=\pi+n\pi/3$, $n=1,2,\ldots$,
the first two discontinuities in $\operatorname*{Im}A$ being $-mR/2 $, next
two are twice of this amount, then next two trice, etc.%

\begin{figure}
[ptb]
\begin{center}
\includegraphics[
height=3.4688in,
width=4.7193in
]%
{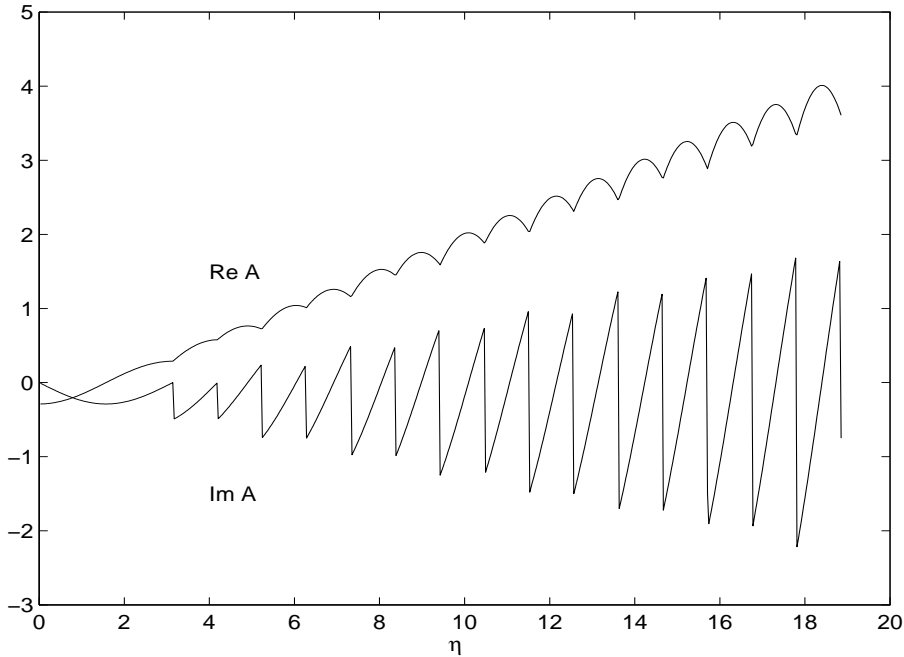}%
\caption{Real and imaginary parts of $A(\eta)/mR$ at $b^{2}=1/2$.}%
\label{fig2}%
\end{center}
\end{figure}

\begin{figure}
[ptb]
\begin{center}
\includegraphics[
height=3.4688in,
width=4.7193in
]%
{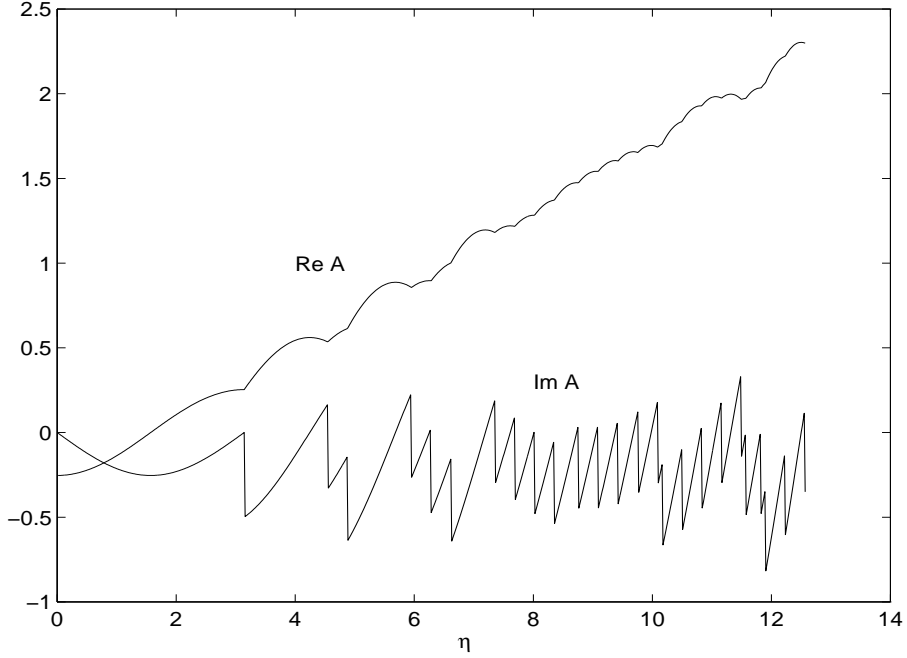}%
\caption{Real and imaginary parts of $A(\eta)/mR$ at $b^{2}=0.8086\ldots$.}%
\label{fig3}%
\end{center}
\end{figure}

With irrational $b^{2}$ the picture is far less regular. To illustrate what
happens when $b^{2}$ slightly deviates from a simple rational number, in fig.3
we plot $A(\eta)$ for $b^{2}=0.8086\ldots$ which is reasonably close to $1$.
Comparing with fig.1 we see that the first discontinuity at $\eta=\pi$ remains
basically the same while the second (double) discontinuity at $\eta=3\pi/2$
splits in two simple ones, the third (triple) splits in three simple jumps
etc. At some point these splitted groups come to overlap and the picture turns irregular.

\section{Staircase situation\label{Staircase}}

The staircase model \cite{staircase} is a formal analytic continuation of ShG
to complex values of the parameter $b$ such that $b^{-1}=b^{*}$. Although the
physical content of this continuation is not completely clear from the field
theory point of view, the TBA equation (\ref{TBA}) remains completely sensible
and this continuation of (\ref{TBA}) can be studied on its own footing. The
effective central charge (\ref{ceff}) is still real and develops sometimes
quite intriguing patterns (see \cite{staircase}).

In \cite{staircase} the complex $b$ has been parameterised as follows
\begin{equation}
b^{2}=\frac{1+2i\theta_{0}/\pi}{1-2i\theta_{0}/\pi}\label{btheta0}%
\end{equation}
with real $0\leq\theta_{0}<\infty$. Parameter $p$ of (\ref{p}) now reads
\begin{equation}
p=\frac12+\frac{i\theta_{0}}\pi\label{ptheta0}%
\end{equation}
while $1-p=p^{*}$ and $a$ defined in (\ref{a}) is purely imaginary
$a=-2i\theta_{0}/\pi$. The TBA kernel (\ref{kernel}) is real and reads
\begin{equation}
\varphi(\theta)=\frac1{2\pi}\left(  \frac1{\cosh(\theta+\theta_{0})}%
+\frac1{\cosh(\theta-\theta_{0})}\right) \label{kertheta0}%
\end{equation}
with the Fourier transform
\begin{equation}
\varphi(\omega)=\frac{\cos(\omega\theta_{0})}{\cosh(\pi\omega/2)}\label{fker0}%
\end{equation}

After these substitution the integral equation (\ref{TBA}) determines
real-analytic functions $\varepsilon(\theta)$, $Y(\theta)$ and, through
(\ref{X}), a real-analytic and symmetric $X(\theta)$ with the asymptotic
behavior at $\operatorname*{Re}\theta\rightarrow\infty$ in the strip
$-\pi/2<\operatorname*{Im}\theta<\pi/2$%
\begin{equation}
X(\theta)\sim\exp\left(  -\frac{mR}{4\cosh\theta_{0}}\exp\theta\right)
\label{Xasymp0}%
\end{equation}
The functional equation (\ref{XX}) reads now
\begin{equation}
X\left(  \theta+i\pi/2\right)  X\left(  \theta-i\pi/2\right)  =1+X\left(
\theta+\theta_{0}\right)  X\left(  \theta-\theta_{0}\right) \label{XX0}%
\end{equation}
All the considerations of sect.\ref{Application} can be repeated literally.
Function $X(\theta)$ is still entire as well as $T(\theta)$ and $\tilde
T(\theta)$ defined in eq.(\ref{shGT}) and the asymptotic (\ref{Xasymp0}) can
be extended to the strip $-\pi<\operatorname*{Im}\theta<\pi$. The difference
is that the periods of $T\left(  \theta\right)  $ and $\tilde T(\theta)$
\begin{align}
T(\theta+\tau)  & =T\left(  \theta\right)  \;;\;\;\;\tau=i\pi(1+a)/2=i\pi
/2+\theta_{0}\label{periods}\\
\tilde T(\theta+\tilde\tau)  & =\tilde T(\theta)\;\;;\;\;\;\tilde\tau
=i\pi(1-a)/2=i\pi/2-\theta_{0}\nonumber
\end{align}
are now complex $\tau=-\tilde\tau^{*}$. Functions $T(\theta)=T\left(
-\theta\right)  $ and $\tilde T(\theta)=\tilde T(-\theta)$ are still symmetric
but no more real analytic. Instead
\begin{equation}
T^{*}(\theta)=\tilde T(\theta^{*})\label{Tconj}%
\end{equation}
Expansions similar to (\ref{Tch})
\begin{align}
T(\theta) &  =\sum_{-\infty}^{\infty}T_{n}\exp(2i\pi n\theta/\tau
)\label{Texp}\\
\tilde T(\theta) &  =\sum_{-\infty}^{\infty}\tilde T_{n}\exp(2i\pi
n\theta/\tilde\tau)\nonumber
\end{align}
are convergent and $T_{n}=T_{-n}$; $\tilde T_{n}=\tilde T_{-n}$. Instead of
being real as in the real $b$ case, these coefficients are complex conjugate
$T_{n}^{*}=\tilde T_{n}$.%

\begin{figure}
[tb]
\begin{center}
\includegraphics[
height=3.4688in,
width=4.7288in
]%
{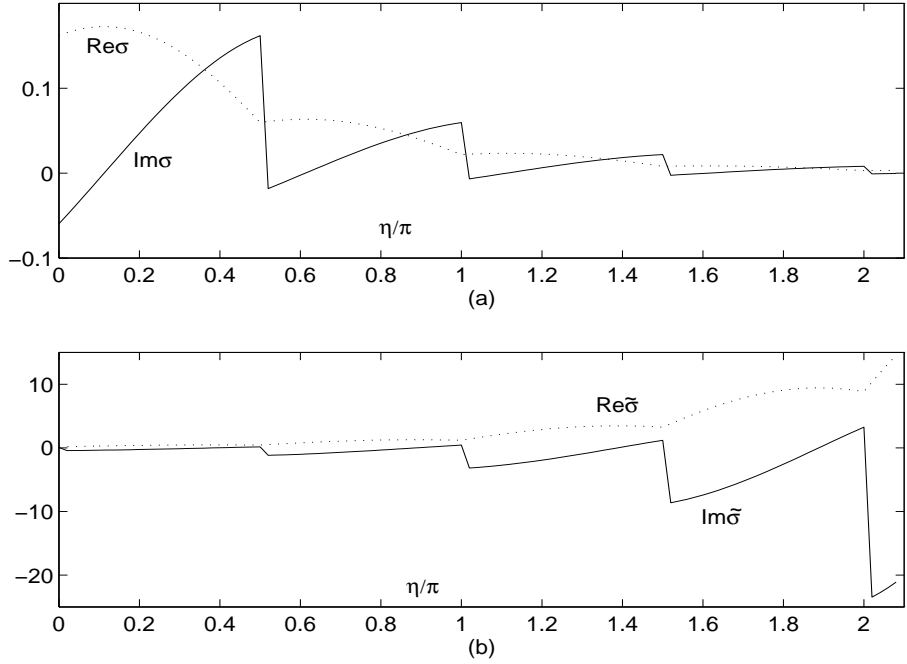}%
\caption{Real and imaginary parts of $\sigma(\eta)/mR$ (a) and $\tilde{\sigma
}(\eta)/mR$ (b). The staircase example for $\theta_{0}=1$.}%
\label{fig10}%
\end{center}
\end{figure}

In the strip $0<\operatorname*{Im}\theta<\pi/2$ the following
$\operatorname*{Re}\theta\rightarrow\infty$ asymptotics holds for $T$ and
$\tilde T$
\begin{align}
T(\theta)  & \sim X(\theta+\tilde\tau)/X(\theta)\sim\exp\left(  \frac
{mR(1-i\exp(-\theta_{0}))}{4\cosh\theta_{0}}e^{\theta}\right) \label{Tasymp0}%
\\
\tilde T(\theta)  & \sim X(\theta+\tau)/X(\theta)\sim\exp\left(
\frac{mR(1-i\exp(\theta_{0}))}{4\cosh\theta_{0}}e^{\theta}\right) \nonumber
\end{align}
The asymptotic changes at the Stokes line along the real axis where
\begin{align}
T(\theta)  & \sim2\exp\left(  \frac{mR(1+i\sinh\theta_{0})}{4\cosh\theta_{0}%
}e^{\theta}\right)  \cos\left(  \frac{mR}4e^{\theta}\right) \label{Tstokes0}\\
\tilde T(\theta)  & \sim2\exp\left(  \frac{mR(1-i\sinh\theta_{0})}%
{4\cosh\theta_{0}}e^{\theta}\right)  \cos\left(  \frac{mR}4e^{\theta}\right)
\nonumber
\end{align}
and we observe again an infinite sequence of zeroes accumulating at infinity,
the density being the same as in the real $b$ case of sect.\ref{LargeRe}. In
the strips $\pi n/2<$Im$\theta<\pi(n+1)/2$ (with arbitrary integer $n$) the
Re$\theta\rightarrow\infty$ asymptotics follow from the periodicity
(\ref{periods}). In particular, along the lines Im$\theta=in\pi/2$ (with any
integer $n$) we'll have
\begin{align}
T(\theta)  & \sim2\exp\left(  \frac{mR(1+i\sinh\theta_{0})}{4\cosh\theta_{0}%
}e^{\operatorname*{Re}\theta-n\theta_{0}}\right)  \cos\left(  \frac
{mR}4e^{\operatorname*{Re}\theta-n\theta_{0}}\right) \label{Tstokesn}\\
\tilde T(\theta)  & \sim2\exp\left(  \frac{mR(1-i\sinh\theta_{0})}%
{4\cosh\theta_{0}}e^{\operatorname*{Re}\theta+n\theta_{0}}\right)  \cos\left(
\frac{mR}4e^{\operatorname*{Re}\theta+n\theta_{0}}\right) \nonumber
\end{align}

Functions $\sigma(\eta)$ and $\tilde\sigma(\eta)$ which control the
asymptotics of $T$ and $\tilde T$ at $\operatorname*{Re}\theta\rightarrow
\infty$%
\begin{align}
T(\theta)  & \sim\exp\left(  \sigma(\operatorname*{Im}\theta
)e^{\operatorname*{Re}\theta}\right) \label{Tsigma}\\
\tilde T(\theta)  & \sim\exp\left(  \tilde\sigma(\operatorname*{Im}%
\theta)e^{\operatorname*{Re}\theta}\right) \nonumber
\end{align}
are plotted in fig.\ref{fig10} for the case $\theta_{0}=1$. It's enough to
present them for $\eta\geq0$ since $\sigma(-\eta)=\tilde\sigma^{*}(\eta)$. The
imaginary part of $\sigma(\eta)$ jumps at the points $\eta=n\pi/2$ by the
amount $-mR\exp(-n\theta_{0})/2$, in accord with the density of zeroes
predicted by (\ref{Tstokesn}). Unlike the previously considered case of real
$b$, in the staircase situation the zeroes of $T$ and $\tilde T$ are not
located exactly at the lines $\operatorname*{Im}\theta=in\pi/2$ but slightly
shifted in the imaginary direction (we'll observe this deviation numerically
in the next section) and approach these lines asymptoticlally as
$\operatorname*{Re}\theta\rightarrow\infty$.

Notice also that e.g. $T(\theta)$ is a single valued function of $\xi
=\exp(2i\pi\theta/\tau)$. In the complex plane of this variable the asymptotic
lines of accumulation of zeroes $\operatorname*{Im}\theta=0$;
$\operatorname*{Re}\theta\rightarrow\pm\infty$ are parts of the spiral
$\left|  \xi\right|  =\exp(\pi\arg\xi/2\theta_{0})$ near which zeroes become
dense at $\left|  \xi\right|  \rightarrow\infty$ or $\left|  \xi\right|
\rightarrow0$, the density growing as $\left|  \xi\right|  ^{\pm
(1/4+\theta_{0}^{2}/\pi^{2})}$ respectively. Therefore the large (or
small)$\left|  \xi\right|  $ asymptotics of $T(\xi)$ at fixed $\arg\xi$ is
rather complicated.%

\begin{figure}
[tb]
\begin{center}
\includegraphics[
height=3.4688in,
width=4.7288in
]%
{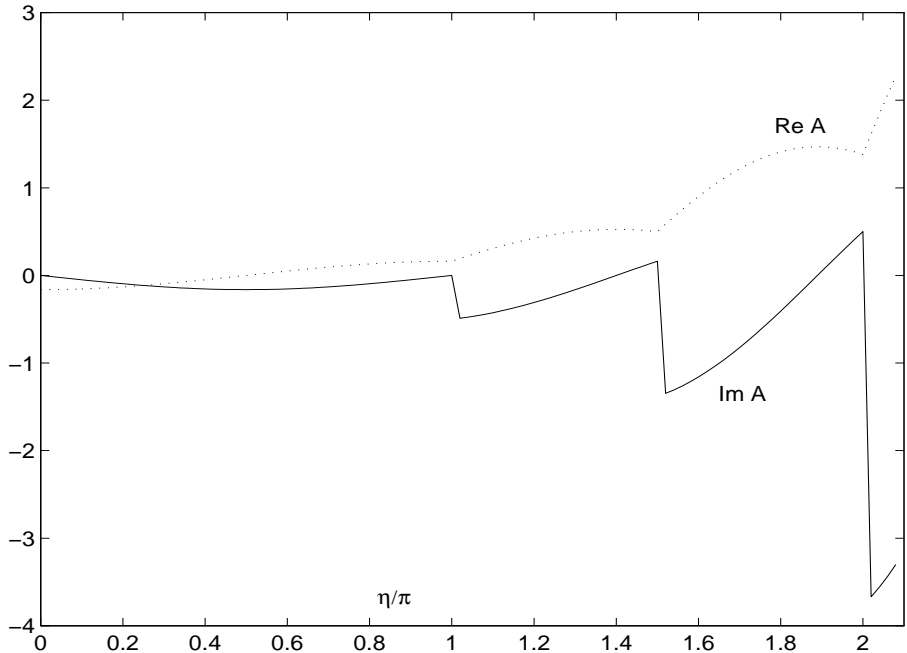}%
\caption{A staircase example of $A(\eta)/mR$ at $\theta_{0}=1$.}%
\label{fig11}%
\end{center}
\end{figure}

Large $\operatorname*{Re}\theta$ asymtotic of $X(\theta)$ at fixed
$\eta=\operatorname*{Im}\theta$ is controlled by the function $A(\eta)$ (see
eq.(\ref{Asymp})). An example corresponding to the case $\theta_{0}=1$ is
presented in fig.\ref{fig11}. At the points $\eta=\pm\pi(n+1)/2$,
$n=1,2,\ldots$ imaginary part of $A(\eta)$ has discontinuities equal to
$-mR\sinh(n\theta_{0})/(2\sinh\theta_{0})$. These amounts determine the
asymptotic density of zeroes of $X(\theta)$ along these lines.

\section{Numerics}

Integral equation (\ref{TBA}) can be easily solved numerically e.g., by
iterations. The iterations happen to be well convergent (the convergence is
somewhat slower if $R$ approaches to $0$ or the parameter $p$ is taken very
small). In the strip $|\operatorname*{Im}\theta|<\pi/2$ function $X(\theta)$
can be computed using the integral representation (\ref{X}). This allows to
continue $X(\theta)$ to the whole complex plane iterating the relation
(\ref{XX}) (in fact at large Im$\theta$ it is more convenient to evaluate
first $T(\theta)$ inside its period and then use (\ref{shGT})). In the rest of
this section we will use the logarithmic scale parameter $x=\log(mR/2)$
instead of $R$.%

\begin{figure}
[ptb]
\begin{center}
\includegraphics[
height=3.4688in,
width=4.7193in
]%
{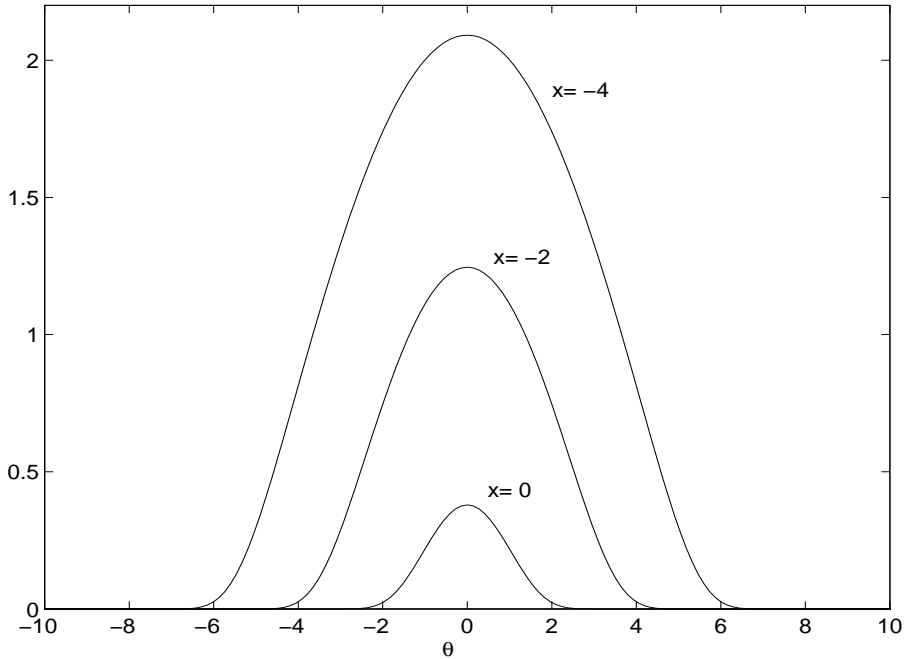}%
\caption{Typical bell-shaped patterns of $X(\theta)$ on the real axis
($b^{2}=1$).}%
\label{fig4}%
\end{center}
\end{figure}

\textbf{1. Self-dual point }$\mathbf{b}^{2}\mathbf{=1}$.

In fig.\ref{fig4} several examples of function $X(\theta)$ on the real axis of
$\theta$ are plotted for different values of $x$. Function is typically
bell-shaped. As $x$ becomes large negative the width of the support of the
bell as well as its height grow proportionally to $-x$. No plateau typical for
perturbed rational CFT's is developed.%

\begin{figure}
[ptb]
\begin{center}
\includegraphics[
height=3.4688in,
width=4.7193in
]%
{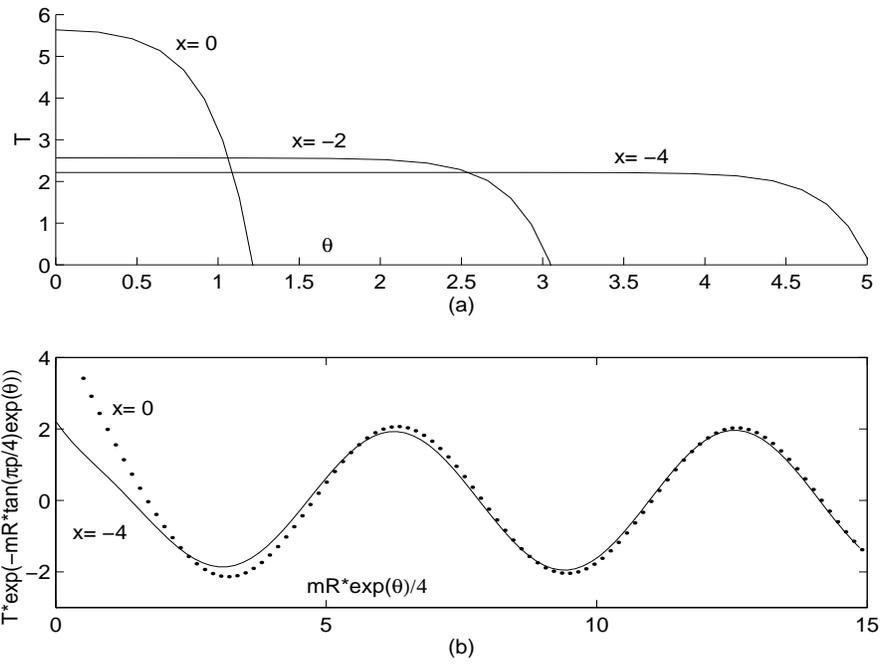}%
\caption{Function $T(\theta)$ on the real axis ($b^{2}=1$).}%
\label{fig5}%
\end{center}
\end{figure}
Few samples of $T(\theta)$ (which is the same as $\tilde T(\theta)$ at the
self-dual point) are presented in fig.\ref{fig5}. At $x$ negative and large
enough, $T$ develops a plateau in the ``central region'' $x<\theta<-x$ of the
height which approaches slowly to $2 $ as $-x$ grows. We'll comment more about
this approach below. Outside the central region it starts to oscillate with
growing amplitude and friquency. Approach to the asymptotic (\ref{Tstokes}) is
very fast.%

\begin{figure}
[ptb]
\begin{center}
\includegraphics[
height=3.4688in,
width=4.7288in
]%
{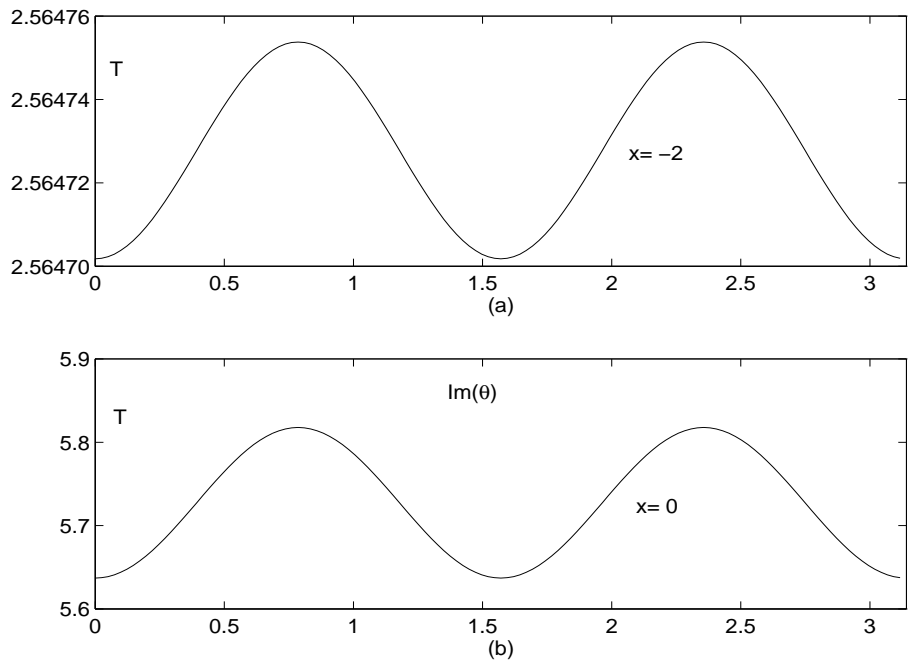}%
\caption{Function $T(\theta)$ along the imaginary axis ($b^{2}=1$).}%
\label{fig7}%
\end{center}
\end{figure}

Due the the symmetry $T(\theta)=T(-\theta)$ this function is real on the
imaginary axis too. A couple of examples are plotted in fig.\ref{fig7}. At $x
$ essentially negative, when the plateau is well developped in the central
region, the mean value $T_{0}$ (see eq.(\ref{Tch})) is very close to the
plateau height, the oscillations around (determined mainly by $T_{1}$) being
very small ($T_{1}\sim R^{4}$, see eq.(\ref{Tnn}) below).%

\begin{figure}
[ptb]
\begin{center}
\includegraphics[
height=3.4688in,
width=4.7288in
]%
{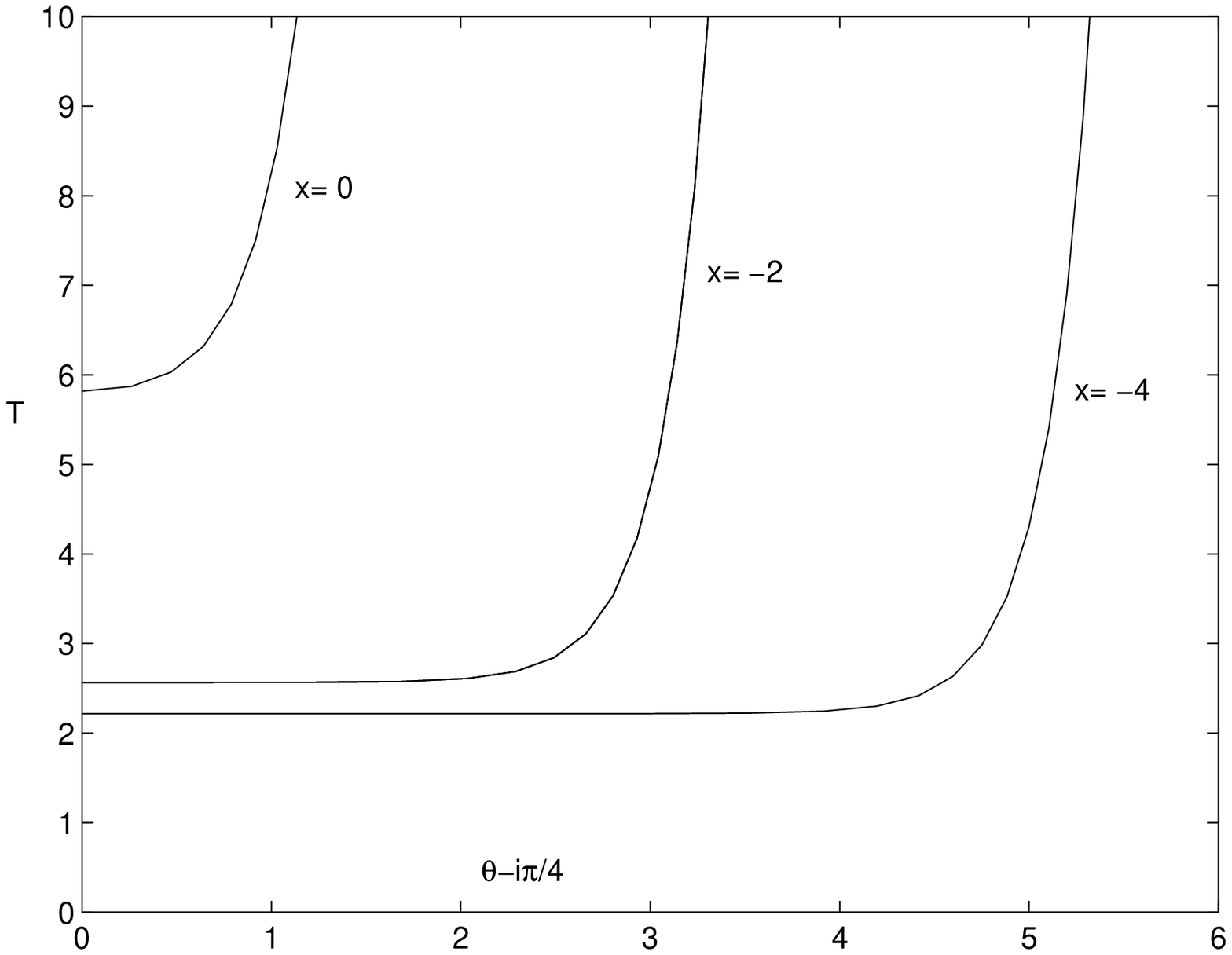}%
\caption{Half-period ($i\pi/4$) shifted $T(\theta)$ at the self-dual point
$b^{2}=1$.}%
\label{fig6}%
\end{center}
\end{figure}

Function $T(\theta)$ is real also at the half-period line $\operatorname*{Im}%
\theta=\pi/4$. Again there is a plateau region (at large negative $x$) of the
same height as on the real axis. Then $T(\theta+i\pi/4)$ remains positive and
grows following the asymptotic (\ref{Thalf}) (see fig.\ref{fig6}). Numerical
comuputations in the whole period strip $0\leq$Im$\theta<\pi/2$ show no sign
of other zeros then those on the real axis.%

\begin{figure}
[ptb]
\begin{center}
\includegraphics[
height=3.4688in,
width=4.7089in
]%
{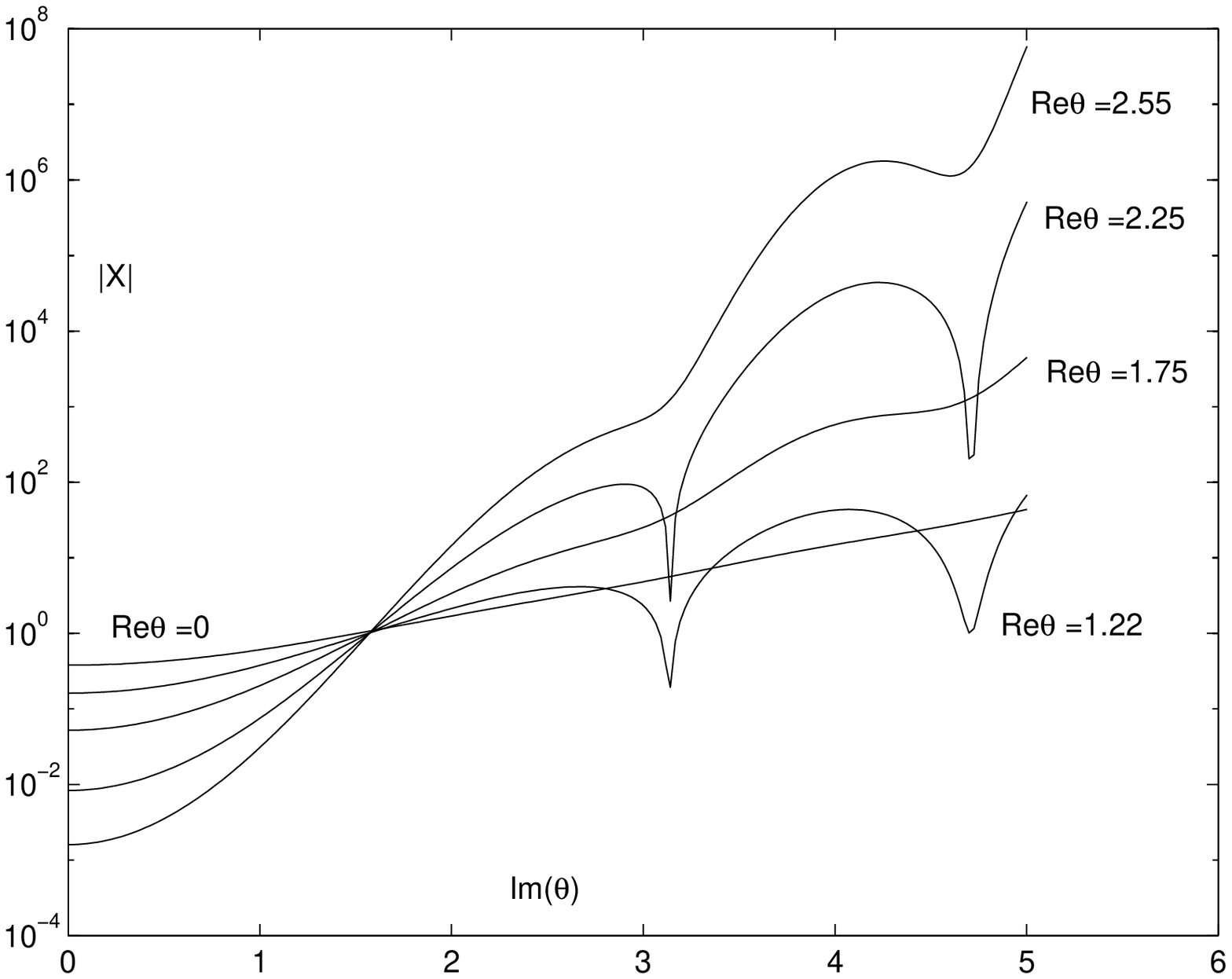}%
\caption{$|X(\theta)|$ as a function of $\operatorname*{Im}\theta$ at
different values of $\operatorname*{Re}\theta$ ($b^{2}=1$ and $x=0$).}%
\label{fig8}%
\end{center}
\end{figure}

\begin{figure}
[ptb]
\begin{center}
\includegraphics[
height=3.4688in,
width=4.7288in
]%
{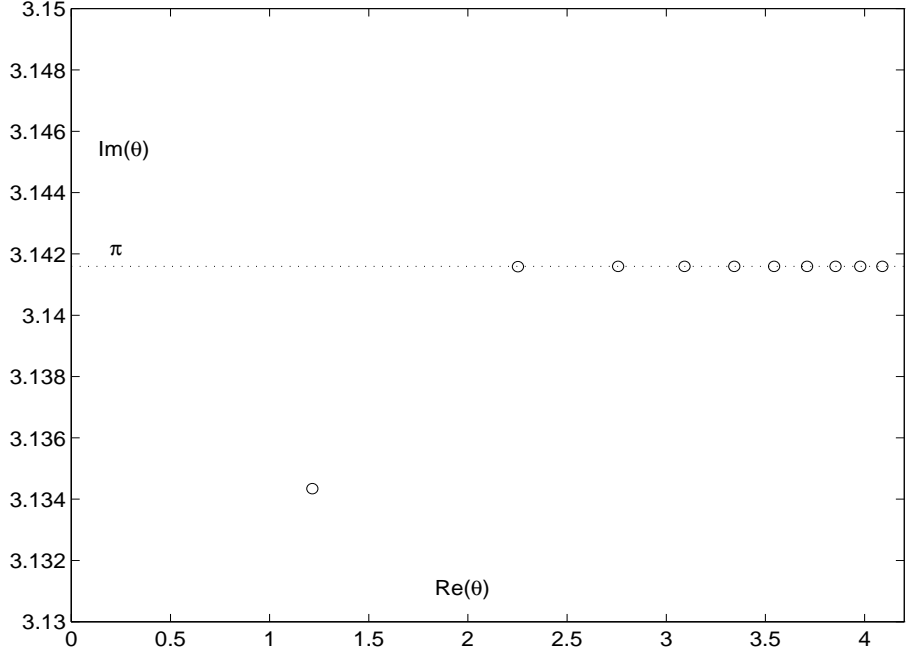}%
\caption{First 10 zeros of $X(\theta)$ located near the line
$\operatorname*{Im}\theta=\pi$. An example for $x=0$ and $b^{2}=1$.}%
\label{fig9}%
\end{center}
\end{figure}

In fig.\ref{fig8} few examples of $\left|  X(\theta)\right|  $ in the complex
plane at different values of $\operatorname*{Re}\theta$ are plotted vs.
$\operatorname*{Im}\theta$ (for $x=0$). The specific values
$\operatorname*{Re}\theta=1.22$ and $\operatorname*{Re}\theta=2.25$ are chosen
close to the positions of the first two zeros of $T(\theta)$ on the real axis.
The deeps near $\operatorname*{Im}\theta=\pi$ and $\operatorname*{Im}%
\theta=3\pi/2$ indicate a presence of zeros of $X(\theta)$ nearby.

More precise positions of zeros of $X(\theta)$ near the line
$\operatorname*{Im}\theta=\pi$ (for the same value $x=0$) are examplified in
fig.\ref{fig9}. In fact all these zeros are inside the strip
$|\operatorname*{Im}\theta|<\pi$. Only the first zero deviates noticably from
the line $\operatorname*{Im}\theta=\pi$. The imaginary parts of next zeros are
already very close to $\pi$ and tend to this value very fast.

\textbf{2. Rational points}. As an example of a rational point we take the
simplest case $b^{2}=1/2$. The periods of $T(\theta)$ and $\tilde{T}(\theta)$
are commensurable and equal to $2i\pi/3$ and $i\pi/3$ respectively. In fact in
this case there is no need to study separately $T$ and $\tilde{T}$ since, as
it is readily derived from (\ref{XX}), they are bound up by the relation
\begin{equation}
\tilde{T}(\theta)=T(\theta)T(\theta+i\pi/3)-2\label{T2}%
\end{equation}
It should be noted that similar finite degree functional relations between $T
$ and $\tilde{T}$ exist for any rational $b^{2}$. For example, at $b^{2}=1/3$
the periods of $T$ and $\tilde{T}$ are $3i\pi/4$ and $i\pi/4$ respectively
and
\begin{equation}
\tilde{T}(\theta)=T(\theta)T(\theta+i\pi/4)T(\theta+i\pi/2)-T(\theta
)-T(\theta+i\pi/4)-T(\theta+i\pi/2)
\end{equation}%
\begin{figure}
[tb]
\begin{center}
\includegraphics[
height=4.0093in,
width=4.7288in
]%
{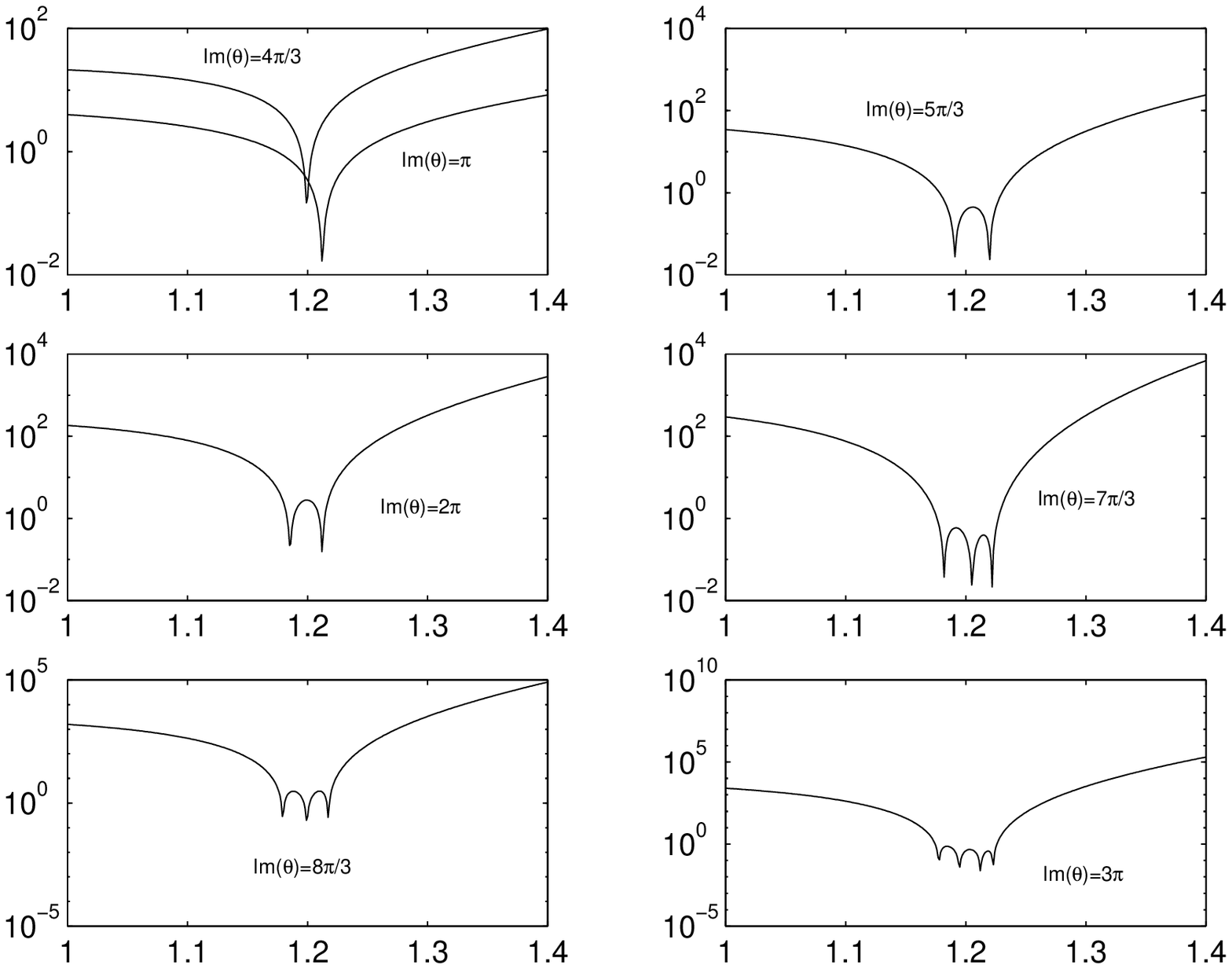}%
\caption{$|X(\theta)|$ vs. $\operatorname*{Re}\theta$ in the vicinity of the
first (multi-)zero at $\operatorname*{Re}\theta\approx1.2$ and different
$\operatorname*{Im}\theta$ ($x=0$, $b^{2}=1/2$).}%
\label{fig13}%
\end{center}
\end{figure}

\begin{figure}
[ptbh]
\begin{center}
\includegraphics[
height=3.2993in,
width=4.529in
]%
{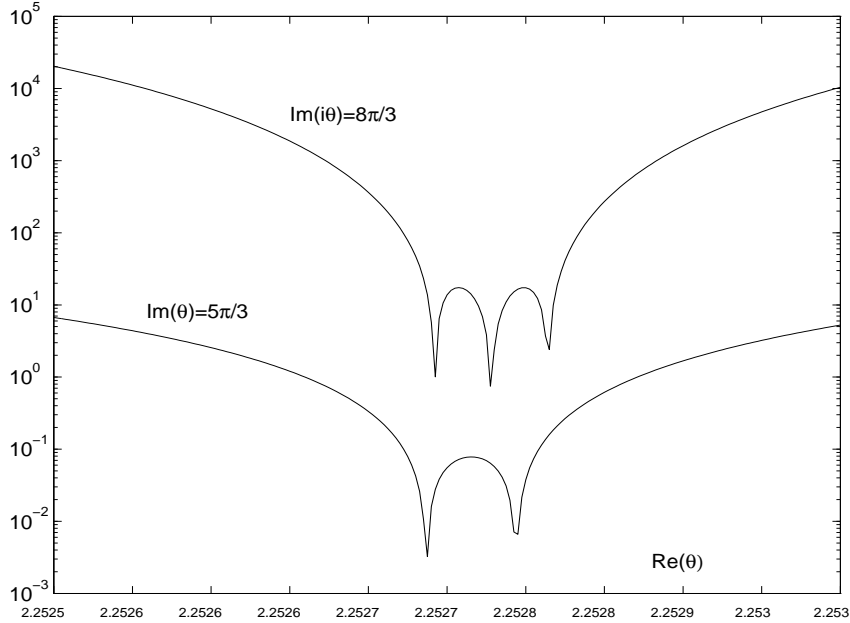}%
\caption{The same as in fig.12 near the next multi-zero at $\operatorname*{Re}%
\theta\approx2.2527\ldots$.}%
\label{fig14}%
\end{center}
\end{figure}

Numerical patterns of $T\left(  \theta\right)  $, $\tilde T(\theta)$ and
$X(\theta)$ are essentially the same as for $b^{2}=1$. I'd only show few plots
of $|X(\theta)|$ in the complex plane along the lines Im$\theta=\pi$, $4\pi
/3$, $5\pi/3$, $2\pi$, $7\pi/3$ etc. to illustrate the mechanism of
multiplication of the zeros density in the asymptotics $\operatorname*{Re}%
\theta\rightarrow\infty$ as required by the prediction of fig.\ref{fig2}. In
fig.\ref{fig13} $|X(\theta)|$ is plotted at $\operatorname*{Re}\theta$ in the
vicinity of the real position of the first zero in $T(\theta)$ at
$\theta=1.2241\ldots$ (the case $x=0$ is taken as an example). At
$\operatorname*{Im}\theta=\pi$ and $4\pi/3$ the picture indicates simple zeros
located closely to this point in $\operatorname*{Re}\theta$ and slightly
displaced in the imaginary direction. At $\operatorname*{Im}\theta=5\pi/3$ and
$2\pi$ the zeros are splitted in two closely located ones again near the same
position in the real direction. For $\operatorname*{Im}\theta=7\pi/3$ and
$8\pi/3$ there are triplets of close zeros, etc. In fig.\ref{fig14} the same
is examplified near the next zero of $T(\theta)$ at $\theta=2.2527\ldots$. It
is seen already that the scale of splitting becomes very small with
$\operatorname*{Re}\theta$ growing and such zero multiplets look like multiple
zeros if the numerical resolution is not enough.%

\textbf{3. General real }$\mathbf{b}^{2}$. In general the periods of
$T(\theta)$ and $\tilde T(\theta)$ are incommensurable. As it was mentioned in
sect.\ref{LargeRe}, it leads in particular to quite complicated
$\operatorname*{Re}\theta\rightarrow\infty$ asymptotics at sufficiently large
$\operatorname*{Im}\theta$. While the analytic structure of $T(\theta)$ and
$\tilde T(\theta) $ remains essentially as described above (in particular, I
verified for many examples that all zeroes of $T$ and $\tilde T$ are on the
real axis), the structure of zeros in $X(\theta)$ becomes, as
$\operatorname*{Im}\theta$ comes essentially large, rather chaotic. I hope to
turn again to this point in future studies. Let me mention only the following
observation conserning the small $R$ (or large negative $x$) picture. If
$-x\gg1$, in the central region $x<\theta<-x$ function $X(\theta)$ matches
extremely well the following expression
\begin{equation}
X(\theta)=\frac{\cos(2QP\theta)}{\left[  \sinh(2\pi bP)\sinh(2\pi P/b)\right]
^{1/2}}\label{Xcos}%
\end{equation}
where $P$ is an $R$-dependent parameter. Roughly it can be estimated from the
requirement that $X(\theta)\simeq0$ at $\theta=\pm x$, i.e., $P=\pi
/(-4Qx)+O(x^{-2})$. Notice that substituting of this approximation to the
expression of the effective central charge
\begin{equation}
c_{\text{eff}}=1-24P^{2}\label{ceffP}%
\end{equation}
(see ref.\cite{Liouville} for the motivations) we arrive just at the leading
UV logarithmic correction (\ref{Eirreg}). It is easy to verify that expression
(\ref{Xcos}) \emph{satisfies exactly} the functional $X$-system (\ref{XX}).
This means in particular that it remains a valid approximation (at large $-x$)
of $X(\theta)$ in the whole complex strip $x<\;\operatorname*{Re}\theta<-x$.%

\begin{figure}
[ptbh]
\begin{center}
\includegraphics[
height=3.4688in,
width=4.7288in
]%
{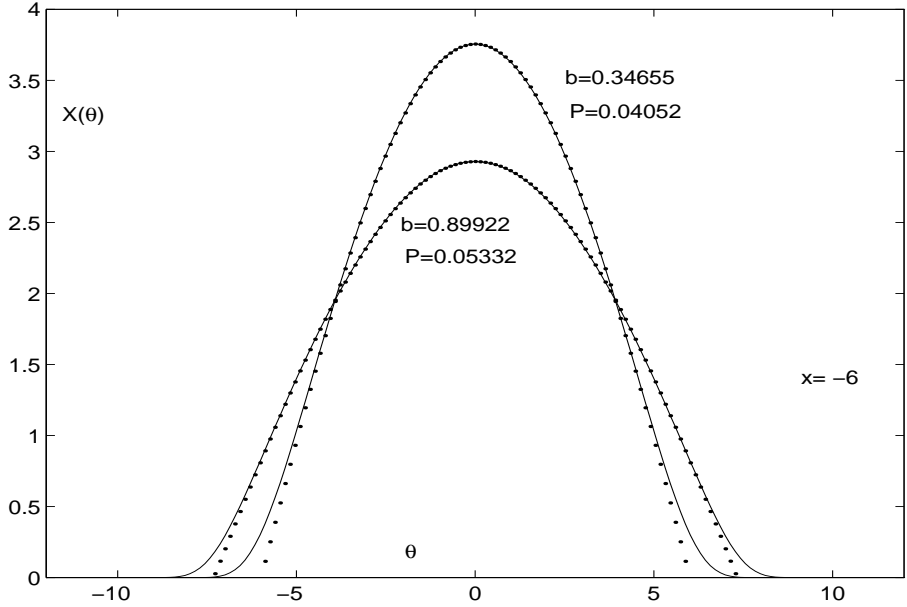}%
\caption{Function $X(\theta)$ at $x=-6$ and different values of $b^{2}$ (solid
lines) compared with the approximation (7.3). Corresponding values of $P$ are
indicated. }%
\label{fig15}%
\end{center}
\end{figure}

In fact along the lines of \cite{Liouville} and \cite{Ahn} a far better
estimate of $P$ can be found which takes into account all logarithmic in $R$
corrections to (\ref{Eirreg}). In this framework $P$ is determined as the
first root of the transcendental equation
\begin{equation}
4QP\log\left(  R/2\pi\right)  +i\log(-S_{L}(P))=-\pi\label{LQC}%
\end{equation}
where $S_{L}(P)$ is the so-called Liouville reflection amplitude (for the
arguments see \cite{Liouville})
\begin{equation}
S_{L}(P)=-\left(  \pi\mu\frac{\Gamma(b^{2})}{\Gamma(1-b^{2})}\right)
^{-2iP/b}\frac{\Gamma(1+2ibP)\Gamma(1+2iP/b)}{\Gamma(1-2ibP)\Gamma
(1-2iP/b)}\label{SL}%
\end{equation}
For expample, in fig.\ref{fig15} the shape of $X(\theta)$ is compared with the
approximation (\ref{Xcos}) for $x=-6$ and two values of the parameter
$b^{2}=0.8086\ldots$ and $b^{2}=0.1201\ldots$ According to (\ref{SL}) they
correspond to $P=0.05332\ldots$ and $P=0.04052\ldots$ respectively.

In view of (\ref{Xcos}) the plateau heights of $T(\theta)$ and $\tilde
T(\theta)$ in the central region $x<\theta<-x$ and $x\rightarrow-\infty$ can
be estimated as
\begin{align}
T_{0}  & =2\cosh(2\pi bP)\label{Tcosh}\\
\tilde T_{0}  & =2\cosh(2\pi P/b)\nonumber
\end{align}
with the same $P$ determined by (\ref{SL}).

\textbf{4. Complex (staircase) values of }$\mathbf{b}^{2}$. The staircase
version of TBA equation (\ref{TBA}) (with the kernel (\ref{kertheta0})) is
solved numericaly in the same way as in the ShG case. The structure of the
solution has some interesting differences from the case of real $b$. The
peculiarities are more manifested if parameter $\theta_{0}$ in
(\ref{kertheta0}) is taken sufficiently large and the deep UV region $-x\gg1$
is considered.. At $\theta_{0}\gg1$ parameter $b$ is close to imaginary unity
and
\begin{equation}
Q=\frac\pi{\sqrt{\theta_{0}^{2}+\pi^{2}/4}}\label{Qtheta0}%
\end{equation}
is small.%

\begin{figure}
[ptb]
\begin{center}
\includegraphics[
height=3.4688in,
width=4.7288in
]%
{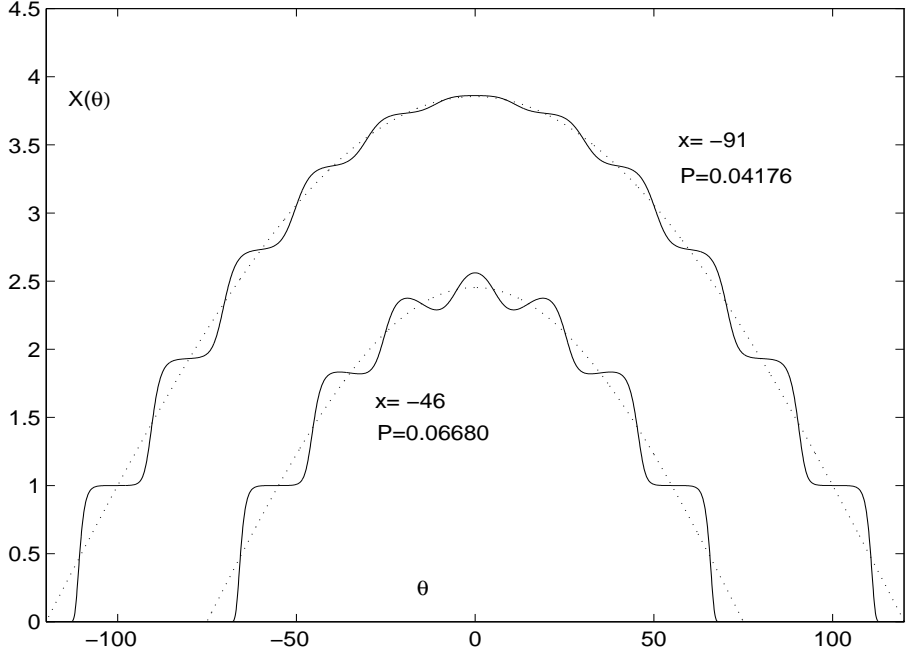}%
\caption{Staircase function $X(\theta)$ at real $\theta$ compared with the
approximation (7.3) ($\theta_{0}=20$).}%
\label{fig16}%
\end{center}
\end{figure}

In fig.\ref{fig16} two pictures of $X(\theta)$ on the real axis are plotted
for $x=-46$ and $x=-91$, both at $\theta_{0}=20$. The characteristic staircase
behavior (in fact very similar to that of $Y(\theta)$ observed in
ref.\cite{staircase}) is very apparent. It is seen also that expression
(\ref{Xcos}) with the parameter $P$ determined as the solution to the
staircase version of (\ref{LQC}) (in the present case $P=0.06680\ldots$ and
$P=0.04176\ldots$ respectively) still follows in the central region the
average behavior of the solution. The deviations (or in other words the
corrections to the approximation (\ref{Xcos})) are now oscillating and much
bigger then in the ShG case. At sufficiently large $-x$ (many amounts of
$\theta_{0}$) the ascending part (starting from $\theta=x-\theta_{0}$) of the
staircase consists of a succession of almost flat steps of constant width
$\theta_{0}$, the heights being very well fitted by the expression
\begin{equation}
X_{n}=\frac{\sin(2QPn/\theta_{0})}{\sin(2QP/\theta_{0})}\label{Xn}%
\end{equation}
with $n=0,1,2,\ldots$. It should be noted that for the reasons to be explained
just below, in the case of complex $b$ expression (\ref{Xcos}) does not give
an approximation in the whole strip $x<\operatorname*{Re}\theta<-x$. Although
(\ref{Xcos}) is still an exact solution to the staircase $X$-system
(\ref{XX0}), its validity is restricted to a certain parallelogram in the
complex $\theta$-plane.%

\begin{figure}
[ptb]
\begin{center}
\includegraphics[
height=3.4688in,
width=4.7288in
]%
{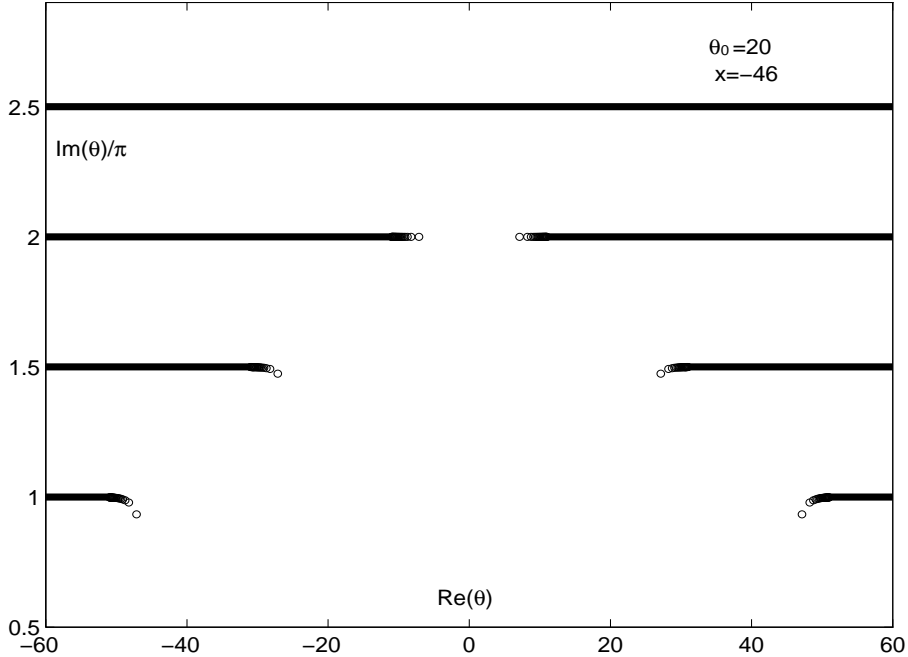}%
\caption{Zeros of $X(\theta)$ at $\theta_{0}=20$ and $x=-46$. Solid lines
stand for very dense sets of zeros.}%
\label{fig17}%
\end{center}
\end{figure}

\begin{figure}
[ptb]
\begin{center}
\includegraphics[
height=3.4688in,
width=4.7288in
]%
{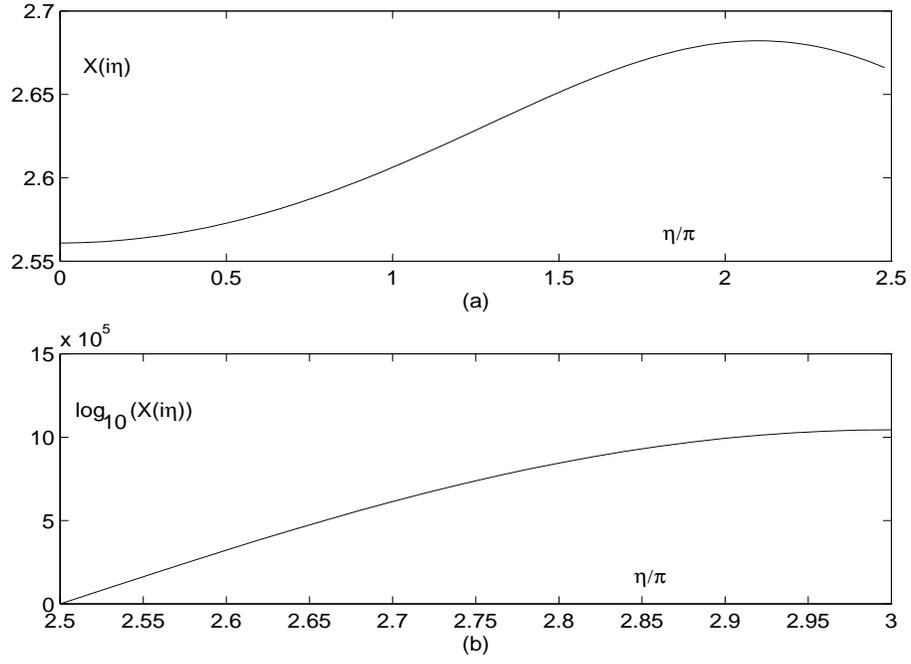}%
\caption{$X(\theta)$ along the imaginary axis ($\theta_{0}=20$, $x=-46$). Plot
(b) is in the decilogarithmic scale.}%
\label{fig18}%
\end{center}
\end{figure}

To see this in fig.\ref{fig17} I present the location of zeros of the function
$X(\theta)$ (at $\theta_{0}=20$ and $x=-46$) in the upper half-plane. The
picture is obviously reflected to the lower half-plane by the symmetry of
$X(\theta)$. We see several strings of zeros accumulating along the lines
$\operatorname*{Im}\theta=\pi n/2$, $n=2,3,4,\ldots$ which start at the values
$\operatorname*{Re}\theta_{2}=47.1500...$, $\operatorname*{Re}\theta
_{3}=27.12267\ldots\approx$ $\operatorname*{Re}\theta_{2}-\theta_{0}$,
$\operatorname*{Re}\theta_{4}=7.11742\ldots\approx$ $\operatorname*{Re}%
\theta_{3}-\theta_{0}$ etc. There are opposite sets of strings symmetric with
respect of the imaginary axis. As $n$ grows they are shifted in steps of
$-\theta_{0}$ and $\theta_{0}$ respectively and meet each other at a certain
value of $n$ ($n=5$ in the present example). After that, continuous lines of
dense zeroes extending from $-\infty$ to $\infty$ are formed. In general at
large $\theta_{0}$ the first such line is already dense enough to produce the
effect of a ``cut'' where the behavior of $X(\theta)$ changes drastically. In
our example, across the first ``cut'' at $\operatorname*{Im}\theta=5\pi/2$ the
absolute value of $X(\theta)$ jumps by many ($10^{5}$) orders of magnitude.
This is examplified in fig.\ref{fig18} where $X(\theta)$ is plotted along the
imaginary axis. Notice that before $\operatorname*{Im}\theta=5\pi/2$ function
$X(\theta)$ is almost constant. As $\theta_{0}$ grows this effect becomes more
and more dramatic and in the limit $\theta_{0}\rightarrow\infty$ the lines of
zeros become real cuts. In the forthcoming paper \cite{msinh} I'm going to
comment more on this effect which plays a crucial role in the analytic
connection between the staircase behavior at finite $\theta_{0}$ and the
``sin-Gordon'' solutions corresponding to purely imaginary $b=i\beta$.%

\begin{figure}
[ptb]
\begin{center}
\includegraphics[
height=3.4688in,
width=4.7288in
]%
{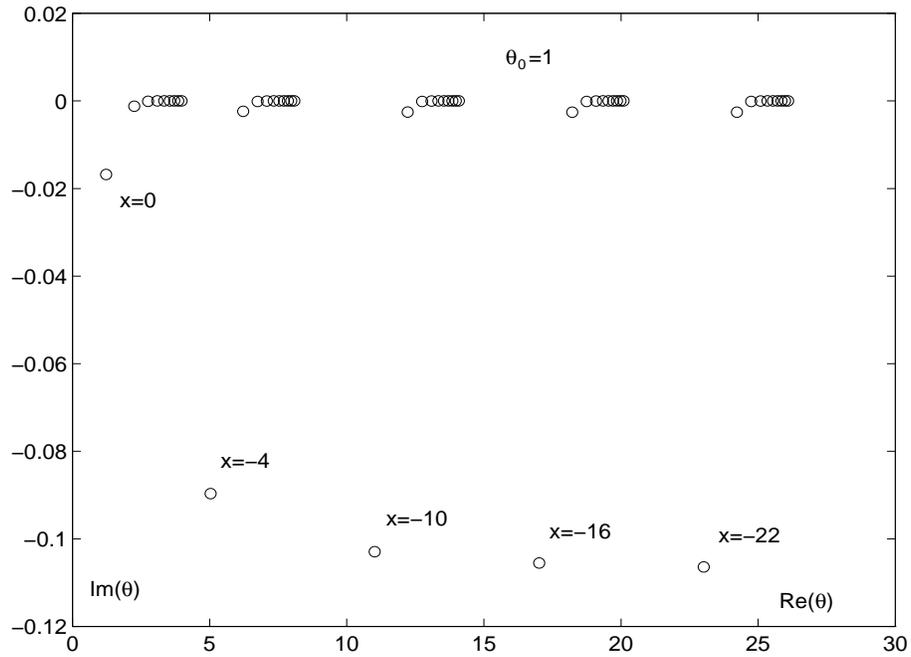}%
\caption{Zeros of $T(\theta)$ at $\theta_{0}=1$ and different $x$.}%
\label{fig12}%
\end{center}
\end{figure}

In fig.\ref{fig12} location of zeros of the periodic function $T(\theta)$ is
shown at the ``moderate'' value $\theta_{0}=1$ and different $x$. The picture
is presented here just to demonstrate two observations: (a) At staircase
values of $b$ zeros of $T(\theta)$ are not exactly on the real axis but
displaced slightly to the complex plane, the displacement becoming negligible
very quickly with the number of zero. (b) At $x\rightarrow-\infty$ the picture
of zeros is ``frozen'' in the sense that at $x$ sufficiently large the pattern
of zeros is simply shifted by the amount $\Delta x$ in the $\theta$ plane as
one changes $x\rightarrow x-\Delta x$. As it is usual in the TBA practice, it
is convenient to study these frosen limiting patterns substituting the
original ``massive'' TBA equation (\ref{TBA}) by the massless version of it.
This is one of the motivations for the subsequent study (to be published
\cite{msinh}).

\section{Concluding remarks}

\begin{itemize}
\item  In the present study I didn't touch at all the important question about
the $R$ (or $x$) dependence of the effective central charge (\ref{ceff})
determined through the finite-size ground state energy (\ref{E0}). The UV
behavior at $x\rightarrow-\infty$ is especially interesting since the analytic
structure of $c_{\text{eff}}(R)$ is quite unusual (like in (\ref{Eirreg})).
The Liouville quantization condition (\ref{LQC}) together with (\ref{ceffP})
proves to be a very good approximation to the UV effective central charge
behavior. However, while it takes into account all the ``soft'' (logarithmic
in $R$) contrubutions to the asymptotic, there are certainly power-like
corrections in $R$. The most important of them (at least at real $b$) is the
contribution of the ground state energy (\ref{Evac}). Approximation
(\ref{ceffP}) is essentially impruved if it is taken into account
\begin{equation}
c_{\text{eff}}=1-24P^{2}+3(mR)^{2}/(4\pi\sin\pi p)\label{ceffPE}%
\end{equation}
Usually after the ground state energy contibution is subtracted the reminder
is a series in the in the perturbative powers of $R$ like in (\ref{Ereg}). In
our present case with Lagrangian (\ref{shg}) naively one could expect a series
like
\begin{equation}
c_{\text{eff}}-3(mR)^{2}/(4\pi\sin\pi p)=1-24P^{2}+\sum_{n=1}^{\infty}%
c_{n}(P)\left(  \mu R^{2+2b^{2}}\right)  ^{2n}\label{cnaive}%
\end{equation}
where the powers of $R$ are predicted on the dimensional arguments and the
coefficients $c_{n}(P)$ are something like the Coulomb gaz perturbative
integrals corresponding to expansion in $\mu$ around the vacuum of momentum
$P$ (therefore they keep some smooth $R$ dependence). Although the leading
correction $n=1$ of order $R^{4+4b^{2}}$ is likely in agreement with the
numerical data (at least at sufficiently small $b^{2}$), the whole structure
(\ref{cnaive}) certainly contradicts duality. Correct expansion must contain
also the ``dual'' corrections with powers $R^{4+4/b^{2}}$. Of course many
people immediately propose to add the dual interaction to the Lagrangian and
consider an action like
\begin{equation}
A_{\mathrm{trivial}}=\int\left[  \frac{1}{4\pi}(\partial_{a}\phi)^{2}%
+2\mu\cosh(2b\phi)+2\tilde{\mu}\cosh(2b^{-1}\phi)\right]  d^{2}x\label{Anaive}%
\end{equation}
with $\tilde{\mu}$ taken from (\ref{mutwiddle}) or sometimes introduced as an
independent coupling. Simultaneous expansion in both couplings would
supposedly lead to a self-dual series
\begin{equation}
c_{\text{eff}}-3(mR)^{2}/(4\pi\sin\pi p)=1-24P^{2}+\sum_{m,n=1}^{\infty
}c_{m,n}(P)\mu^{2m}\tilde{\mu}^{2n}R^{4Q(mb+n/b)}\label{cdouble}%
\end{equation}
Coefficients $c_{m,n}(P)$ are computed as the mixed Coulomb gaz integrals
which incude both kinds of charges produced by the expansion of action
(\ref{Anaive}). Although the general structure of (\ref{cdouble}) looks very
likely, to my sense this scenario (which I call trivial) with naive addition
of dual interactions (as in (\ref{Anaive})) is not exactly in the spirit of
duality. However at present it does not contradict any data and in fact should
be verified. A check of this trivial scenario (as well as any other one)
requires very subtle measurements of the subleading power-like corrections to
the effective central charge as well as tedious calculations of mixed
perturbative integrals. I understand that this is a \emph{quantitative }work
which hardly can be replaced by general speculations.

\item  Of course the definition of $P$ as the solution to the quantization
condition (\ref{LQC}) is not completely unambiguous. The power-like in $R$
corrections (which are exponentially small in $1/P$) can be arbitrarily
redistributed between the expression for the observable effective central
charge (\ref{cdouble}) and the formulation of the quantization condition. In
other words one can add exponentially small in $1/P$ corrections to (\ref{SL})
and consider this as a new definition of $P$. The problem is that at present
the parameter $P$ is not precisely observable, i.e., it cannot be directly
measured in the TBA calculations (apart from the abovementioned
\emph{definition} through the observable finite-size effective central charge).

\item  At this point we arrive at the most intriguing question touched only
slightly in the present study. This is about the possibility to construct the
solutions $Q_{\pm}(\theta)$ and $\tilde{Q}_{\pm}(\theta)$ of eqs.(\ref{QQTQ})
with the properties (\ref{Bloch}) and (\ref{qW1}) such that the solution to
(\ref{TBA}) can be built as the combination (\ref{XQQ1}). Was this be possible
we'd have another unambiguous definition of the parameter $P $ as the Floquet
index in (\ref{Bloch}). However, there are serious doubts that such solutions
exist in any sense, at least for rational values of $b^{2}$. To clarify this
point, in the next publication \cite{msinh} I'll consider the massless version
of the ShG TBA equation where the parameter $P$ in introduced from the very
beginning instead of the scale parameter $R$. In this context the solutions
(\ref{Bloch}) can be found at least as formal series for irrational values of
$b^{2}$ in the way that the construction (\ref{XQQ1}) can be given an exact sense.

\item  In connection with the periodic structures encoded in the functions
$T(\theta)$ and $\tilde{T}(\theta)$, it seems quite interesting to understand
better the $R$ dependence of the coefficients $T_{n}$ and $\tilde{T}_{n}$ in
the expansions (\ref{Tch}) or (\ref{Texp}). I checked numerically the
$R\rightarrow0$ asymptotic of $T_{0}$ and $\tilde{T}_{0}$. The leading
asymptotics of $T_{0}$ and $\tilde{T}_{0}$ (remember that for definiteness in
this study I always take $b\leq1$, in particular the data discussed in this
item were calculated at $b=0.3466$) are very well fitted by the expressions
(\ref{Tcosh}) with $P$ the solution to (\ref{LQC}). The correction to
(\ref{Tcosh}) for $T_{0}$ can be set in the form
\begin{equation}
T_{0}=2\cosh(2\pi bP)+(mR/2\pi)^{4bQ}T_{0}^{(1)}(P)\label{T0corr}%
\end{equation}
where $T_{0}^{(1)}(P)$ is a smoothly varying function of $P$ with certain UV
limit $T_{0}^{(1)}(0)$. As for $\tilde{T}_{0}$ the correction is better fitted
as
\begin{equation}
\tilde{T}_{0}-2\cosh(2\pi P/b)\sim A(P)(mR/2\pi)^{\alpha}\label{TT0corr}%
\end{equation}
with $\alpha$ again numerically close to $4bQ$. The common TBA experience
would expect from the periodic structure of $\tilde{T}(\theta)$ a much smaller
correction $\sim(mR/2\pi)^{4Q/b}$. A probable explanation is that there are
some power-like corrections to the Liouville quantization condition
(\ref{LQC}) so that the ``correct'' value of $P_{\text{correct}}$ is off from
$P$ (calculated from (\ref{LQC})) by an amount of order $(mR/2\pi)^{4bQ}$%
\begin{equation}
P=P_{\text{correct}}+P_{1,0}(mR/2\pi)^{4bQ}+\ldots\label{Pcorr}%
\end{equation}
With this $P_{\text{correct}}$ an asymptotic
\begin{equation}
\tilde{T}_{0}=2\cosh(2\pi P_{\text{correct}}/b)+(mR/2\pi)^{4Q/b}\tilde{T}%
_{0}^{(1)}(P_{\text{correct}})+\ldots\label{tttcorr}%
\end{equation}
must hold. If the power $4Q/b\gg4bQ$ (like in the present experiment with
$4Q/b=37.31\ldots$ $\gg4bQ=4.480\ldots$) we can even try to relate the
coefficient $A$ in (\ref{TT0corr}) to the leading power correction to
(\ref{LQC}). As the corrections in (\ref{tttcorr}) are much more suppressed at
$R\rightarrow0$ this seems reasonable.

\item  It is easy to verify that the leading asymptotic of the higher
coefficients $T_{n}$ and $\tilde T_{n}$ in the expansion (\ref{Tch}) are of
the form
\begin{align}
T_{n}  & =(mR/2\pi)^{2bQ|n|}T_{n}^{(0)}(P)+\ldots\label{Tnn}\\
\tilde T_{n}  & =(mR/2\pi)^{2Q|n|/b}\tilde T_{n}^{(0)}(P)+\ldots\nonumber
\end{align}
with $T_{n}^{(0)}(P)$ and $\tilde T_{n}^{(0)}(P)$ regular at $P=0$. I analysed
quantitatively the functions $T_{1}^{(0)}(P)$ and $T_{1}^{(0)}(P)$. Motivated
by the constructions of refs.\cite{BLZ1,BLZ2} let us introduce slightly
rescaled functions
\begin{align}
t_{1}(P)  & =T_{1}^{(0)}(P)\left(  Z(p)\right)  ^{-2bQ}\label{rescale}\\
\tilde t_{1}(P)  & =\tilde T_{1}^{(0)}(P)\left(  Z(p)\right)  ^{-2Q/b}%
\nonumber
\end{align}
with $Z(p)$ defined in (\ref{Zp}). In table \ref{table1} the values of
$t_{1}(P)$ and $\tilde t_{1}(P)$ are compared with the following analytic
expressions borrowed from refs.\cite{BLZ1,BLZ2} where these coefficients enter
the explicit constructions of the ``sin-Gordon'' (i.e., related to purely
imaginary values of $b$) analogs of $T(\theta)$%
\begin{align}
t_{1}^{\text{CFT}}(P)  & =-\frac{4\pi^{2}\Gamma(1+2b^{2})}{\Gamma^{2}%
(b^{2})\Gamma(1+b^{2}+2ibP)\Gamma(1+b^{2}-2ibP)}\\
\tilde t_{1}^{\text{CFT}}(P)  & =-\frac{4\pi^{2}\Gamma(1+2b^{-2})}{\Gamma
^{2}(b^{-2})\Gamma(1+b^{-2}+2iP/b)\Gamma(1+b^{-2}-2iP/b)}\nonumber
\end{align}
The TBA numbers are measured at $b=0.3465545.\ldots$ and different values of
$x$. The same for the staircase example $\theta_{0}=10$ is presented in
table\ref{table2}. The convergence is noticably slower due to much weeker
suppression of the higher power corrections (Re$(4+4b^{2})=0.1926\ldots$ in
this case). The numbers quoted make it clear that an analytic continuation of
the constructions of refs.\cite{BLZ1,BLZ2} for the ShG or staircase values of
the parameter is quite relevant. The corresponding ``CFT integrable
structures'' will be shown to have a precise relation to the solutions of the
massless versions of ShG and staircase TBA equations \cite{msinh} where the
parameter $P=P_{\text{correct}}$ is fixed by construction.%

\begin{table}[tbp] \centering
\begin{tabular}
[c]{|llllll|}\hline
\multicolumn{1}{|l|}{$x$} & \multicolumn{1}{l|}{$P$} &
\multicolumn{1}{l|}{$t_{1}^{\text{TBA}}$} & \multicolumn{1}{l|}{$t_{1}%
^{\text{CFT}}$} & \multicolumn{1}{l|}{$\tilde t_{1}^{\text{TBA}}$} & $\tilde
t_{1}^{\text{CFT}}$\\\hline
\multicolumn{1}{|l|}{0} & \multicolumn{1}{l|}{0.3896985} &
\multicolumn{1}{l|}{-0.7228948} & \multicolumn{1}{l|}{-0.7213854} &
\multicolumn{1}{l|}{-1.452134e-2} & -1.437650e-2\\
\multicolumn{1}{|l|}{-2} & \multicolumn{1}{l|}{0.1152693} &
\multicolumn{1}{l|}{-0.6585087} & \multicolumn{1}{l|}{-0.6585077} &
\multicolumn{1}{l|}{-8.578078e-3} & -8.578008e-3\\
\multicolumn{1}{|l|}{-4} & \multicolumn{1}{l|}{6.046935e-2} &
\multicolumn{1}{l|}{-0.6542746} & \multicolumn{1}{l|}{-0.6542746} &
\multicolumn{1}{l|}{-8.272562e-3} & -8.272562e-3\\\hline
\end{tabular}
\caption{Numerical values of\label{table1}}%
\end{table}

\begin{table}[tbp] \centering
\begin{tabular}
[c]{|llll|}\hline
\multicolumn{1}{|l|}{$x$} & \multicolumn{1}{l|}{$P$} &
\multicolumn{1}{l|}{$t_{1}^{\text{TBA}}$} & $t_{1}^{\text{CFT}}$\\\hline
\multicolumn{1}{|l|}{0} & \multicolumn{1}{l|}{0.1771863} &
\multicolumn{1}{l|}{0.0711632-1.1669319i} & 0.0051770-1.1896763i\\
\multicolumn{1}{|l|}{-15} & \multicolumn{1}{l|}{8.604775e-2} &
\multicolumn{1}{l|}{-0.0660054-0.7447625i} & -0.0684458-0.7445886i\\
\multicolumn{1}{|l|}{-30} & \multicolumn{1}{l|}{5.695833e-2} &
\multicolumn{1}{l|}{-0.0708064-0.6568688i} & -0.0709316-0.6568002i\\\hline
\end{tabular}
\caption{Numerical values of \label{table2}}%
\end{table}

\item  I'd like to mention a quite intriguing recent article \cite{sinh} where
the author arrived at the function (\ref{X}) in a rather different context. It
appears as the exact wave function of the finite-size sinh-Gordon model in
special ``$\gamma$-coordinates'' which are the sinh-Gordon version of the
Flaschka-McLaughlin variables (see \cite{sinh} for the details). This gives
again a motivation to continue the study of the analytic structures related to
the ShG TBA equation.

\vspace{0.5cm}
\end{itemize}

\textbf{Acknowledgments. }I acknowledge very useful discussions with C.Ahn,
V.Bazhanov, V.Fateev, C.Rim and mostly with A.Zamolodchikov. I thank also
P.Wiegmann who brought my attention to papers \cite{Wiegmann, Zabrodin} and
introduced to me the world of incommensurable periods. Discussions with
G.Mussardo about the ``double sin-Gordon'' \cite{Mussardo} were also quite
relevant. My special gratitude is to S.Lukyanov who communicated me his
article \cite{sinh} before publication and encouraged my work by exciting discussions.

The work is supported in part by EU under the contract ERBFMRX CT 960012.

\end{document}